\documentclass{pasa}
\usepackage[authoryear]{natbib}
\usepackage[figuresleft]{rotating}
\usepackage{graphicx}
\usepackage{caption}
\usepackage{subcaption} 
\usepackage{amssymb,amsmath}
\graphicspath{{./arxiv_graphics/}} 
\newcommand{\offil}[1]{$^{\rm #1}$}
\newcounter{inst} 
\newcommand{\inst}[1]{\noindent%
   \refstepcounter{inst}\offil{\Alph{inst}\label{#1}}  
   }

\newcommand{\farcm}{\mbox{\ensuremath{.\mkern-4mu^\prime}}}
\newcommand{\farcs}{\mbox{\ensuremath{.\!\!^{\prime\prime}}}}
\newcommand{\fdg}{\mbox{\ensuremath{.\!\!^\circ}}}
\newcommand{\arcdeg}{\ensuremath{^{\circ}}}
\newcommand{\arcmin}{\hbox{$^\prime$\,}}
\newcommand{\survarea}{6,100}
\newcommand{\nsrc}{14,110}
\newcommand{\nresolved}{1,247}
\newcommand{\nunresolved}{12,863}
\newcommand{\RArange}{20.5\,h$<$ Right Ascension (RA) $<8.5$\,h}
\newcommand{\Decrange}{$-58^{\circ}<$ Declination (Dec) $<-14^{\circ}$}
\def\fig{Figure}

\def\Fig{Figure}

\def\sect{Section}

\def\Sect{Section}

\def\tab{Table}
\def\tabs{Tables}
\def\Tab{Table}

\def\eqn{equation}

\title[MWACS: A Low-Frequency Catalogue]{The Murchison Widefield Array Commissioning Survey: A
Low-Frequency Catalogue of {\nsrc} Compact Radio Sources over {\survarea} Square Degrees}
\author[Hurley-Walker et al.]{
Natasha Hurley-Walker\offil{\ref{ICRAR}}\thanks{Email: nhw@icrar.org},
John Morgan\offil{\ref{ICRAR}},
Randall B. Wayth\offil{\ref{ICRAR},\ref{CAASTRO}},
Paul J. Hancock\offil{\ref{ICRAR}},
Martin E. Bell\offil{\ref{CAASTRO},\ref{USyd}}, 
Gianni Bernardi\offil{\ref{SKASA},\ref{Rhodes},\ref{CfA}}, 
Ramesh Bhat\offil{\ref{ICRAR}}, 
Frank Briggs\offil{\ref{CAASTRO},\ref{ANU}}, 
Avinash A. Deshpande\offil{\ref{RRI}}, 
Aaron Ewall-Wice\offil{\ref{MIT}}, 
Lu Feng\offil{\ref{MIT}}, 
Bryna J. Hazelton\offil{\ref{UDub}}, 
Luke Hindson\offil{\ref{VUW}},
Daniel C. Jacobs\offil{\ref{ASU}},
David L. Kaplan\offil{\ref{UWM}}
Nadia Kudryavtseva\offil{\ref{ICRAR}}, 
Emil Lenc\offil{\ref{CAASTRO},\ref{USyd}}, 
Benjamin McKinley\offil{\ref{CAASTRO},\ref{ANU}}, 
Daniel Mitchell\offil{\ref{CASS}}, 
Bart Pindor\offil{\ref{CAASTRO},\ref{UMelb}}, 
Pietro Procopio\offil{\ref{CAASTRO},\ref{UMelb}}, 
Divya Oberoi\offil{\ref{NCRA}},
Andr\'{e} Offringa\offil{\ref{CAASTRO},\ref{ANU}}, 
Stephen Ord\offil{\ref{ICRAR},\ref{CAASTRO}},
Jennifer Riding\offil{\ref{CAASTRO},\ref{UMelb}},
Judd D. Bowman\offil{\ref{ASU}},
Roger Cappallo\offil{\ref{Haystack}},
Brian Corey\offil{\ref{Haystack}},
David Emrich\offil{\ref{ICRAR}},
B. M. Gaensler\offil{\ref{CAASTRO},\ref{USyd}},
Robert Goeke\offil{\ref{Haystack}},
Lincoln Greenhill\offil{\ref{CfA}},
Jacqueline Hewitt\offil{\ref{MIT}},
Melanie Johnston-Hollitt\offil{\ref{VUW}},
Justin Kasper\offil{\ref{CfA}},
Eric Kratzenberg\offil{\ref{Haystack}},
Colin Lonsdale\offil{\ref{Haystack}},
Mervyn Lynch\offil{\ref{ICRAR}},
Russell McWhirter\offil{\ref{Haystack}},
Miguel F. Morales\offil{\ref{UDub}},
Edward Morgan\offil{\ref{MIT}},
Thiagaraj Prabu\offil{\ref{RRI}},
Alan Rogers\offil{\ref{Haystack}},
Anish Roshi\offil{\ref{NRAO}},
Udaya Shankar\offil{\ref{RRI}},
K. Srivani\offil{\ref{RRI}},
Ravi Subrahmanyan\offil{\ref{RRI},\ref{CAASTRO}},
Steven Tingay\offil{\ref{ICRAR},\ref{CAASTRO}},
Mark Waterson\offil{\ref{ICRAR},\ref{ANU}},
Rachel Webster\offil{\ref{CAASTRO},\ref{UMelb}},
Alan Whitney\offil{\ref{Haystack}},
Andrew Williams\offil{\ref{ICRAR}},
Chris Williams\offil{\ref{MIT}}
\newline
\affil{\inst{ICRAR}\,International Centre for Radio Astronomy Research (ICRAR), Curtin University, Perth, Australia}
\affil{\inst{CAASTRO}\,ARC Centre of Excellence for All-sky Astrophysics (CAASTRO)}
\affil{\inst{USyd}\,Sydney Institute for Astronomy (SIfA), School of Physics, The University of Sydney, Sydney, NSW 2006, Australia}
\affil{\inst{SKASA}\,SKA SA, 3rd Floor, The Park, Park Road, Pinelands, 7405, South Africa}
\affil{\inst{Rhodes}\,Department of Physics and Electronics, Rhodes University, PO Box 94, Grahamstown, 6140, South Africa}
\affil{\inst{CfA}\,Harvard-Smithsonian Center for Astrophysics, Cambridge, MA, 02138, USA}
\affil{\inst{ANU}\,Research School of Astronomy \& Astrophysics, Mount Stromlo Observatory, Australian National University, Weston Creek ACT 2611, Australia}
\affil{\inst{RRI}\,Raman Research Institute, Bangalore, India}
\affil{\inst{MIT}\,MIT Kavli Institute for Astrophysics and Space Research, Cambridge, MA, USA}
\affil{\inst{UDub}\,Physics Department, University of Washington, Seattle, WA, USA}
\affil{\inst{VUW}\,School of Chemical and Physical Sciences, Victoria University of Wellington, Wellington, New Zealand}
\affil{\inst{ASU}\,School of Earth and Space Exploration, Arizona State University, Tempe, AZ, USA}
\affil{\inst{UWM}\,Physics Department, University of Wisconsin-Milwaukee, Milwaukee, WI, USA}
\affil{\inst{CASS}\,CSIRO Astronomy and Space Science (CASS), PO Box 76, Epping, NSW 1710, Australia}
\affil{\inst{UMelb}\,School of Physics, The University of Melbourne, Melbourne, Australia}
\affil{\inst{NCRA}\,National Centre for Radio Astrophysics, Pune, India}
\affil{\inst{Haystack}\,MIT Haystack Observatory, Westford, MA, USA}
\affil{\inst{NRAO}\,National Radio Astronomy Observatory, Charlottesville, WV, USA}}
\jid{PASA}
\doi{10.1017/pas.\the\year.xxx}
\jyear{\the\year}
\begin{document}
\begin{abstract}
We present the results of an approximately {\survarea} deg$^2$ 104--196\,MHz radio sky survey performed with the Murchison Widefield Array during instrument commissioning between 2012 September and 2012 December: the Murchison Widefield Array Commissioning Survey (MWACS). The data were taken as meridian drift scans with two different 32-antenna sub-arrays that were available during the commissioning period. The survey covers approximately \RArange, \Decrange over three frequency bands centred on 119, 150 and 180\,MHz, with image resolutions of 6--3\,arcmin.
The catalogue has 3-arcmin angular resolution and a typical noise level of 40\,mJy\,beam$^{-1}$, with reduced sensitivity near the field boundaries and bright sources.
We describe the data reduction strategy, based upon mosaiced snapshots, flux density calibration and source-finding method. We present a catalogue of flux density and spectral index measurements for {\nsrc} sources, extracted from the mosaic, {\nresolved} of which are sub-components of complexes of sources.
\end{abstract}
\begin{keywords}
surveys -- radio continuum: general -- techniques: interferometric
\end{keywords}
%
%
\maketitle
\section{Introduction}

Wide-field radio sky surveys yield information for large samples of Galactic and extra-galactic objects, allowing analysis of emission physics, source populations and cosmic evolution of radio sources such as active galactic nuclei, starforming galaxies, pulsars and supernova remnants.
Surveys in new regions of parameter space, such as frequency, are particularly useful for identifying new and unusual objects and adding constraints to emission models.

While radio surveys such as the Sydney University Molonglo Sky Survey \citep[SUMSS;][]{1999AJ....117.1578B,2003MNRAS.342.1117M} at 843\,MHz and the Molonglo Reference Catalogue \citep[MRC;][]{1981MNRAS.194..693L} at 408\,MHz cover the southern sky, there is no deep survey of the southern sky below these frequencies. The Culgoora Circular Array \citep{1995AuJPh..48..143S} measured approximately 1800 high-frequency-selected sources over a Dec range of $-48^\circ$ to $+35^\circ$ at 80~and 160\,MHz with limiting flux density of 4 and 2\,Jy, respectively. However, these were targeted observations rather than a survey.
A recent low-resolution (26\arcmin) survey by the Precision Array for Probing the Epoch of Reionization \citep[PAPER;][]{2011ApJ...734L..34J} detected $\approx500$ sources at 145\,MHz over 4,800\,deg$^2$.

The new generation of aperture-array telescopes such as the Murchison Widefield Array \citep[MWA;][]{2013PASA...30....7T, 2009IEEEP..97.1497L} and the Low-Frequency Array \citep[LOFAR;][]{2013A&A...556A...2V} offer a chance to conduct all-sky low-frequency surveys at higher angular resolution, sensitivity, and speed than ever achieved before.
Observations using the MWA 32-element prototype array such as \citet{2012ApJ...755...47W} and \citet{2013ApJ...771..105B} have demonstrated the scientific potential of low-frequency observations with the MWA design on a radio-quiet southern site.

The MWA is the only low-frequency precursor\footnote{Defined as a facility exploring SKA technology, science and/or operations on an SKA candidate site: https://www.skatelescope.org/technology/precursors-pathfinders-design-studies/} for the Square Kilometre Array (SKA) and the first of the three SKA precursors to be operational for science. The MWA is located at the Murchison Radio-astronomy Observatory in outback Western Australia, the planned site of the future multi-million element SKA-low array. Thus the MWA explores the characteristics of the site at low frequencies in the pursuit of challenging science, exercising much of the physical infrastructure that will be applied to the SKA over a geographic distance of 800\,km (on-site infrastructure, long haul data transport, data archive infrastructure), and provides a valuable base for SKA verification systems.

This paper presents the MWA Commissioning Survey (MWACS), work undertaken during the commissioning period of the full MWA, using a subset of its capabilities.
The survey aimed to verify instrumental performance, verify our understanding of the MWA primary beam, motivate development of new data processing techniques and create an initial sky model of brighter radio sources at MWA frequencies.
A comprehensive sky model, in particular, is a pre-requisite for calibrating the full MWA, due to its huge field-of-view \citep{2008ISTSP...2..707M}. The observed field also includes two of the three regions chosen for deep ($\approx1000$-hour) integrations for the MWA's Epoch of Reionisation key science program (see \citealt{2013PASA...30...31B} for a summary of the MWA's key science programs). MWACS is the first large-scale survey performed by the MWA in its role as an SKA precursor and the first such survey by an SKA precursor.

These observations cover approximately {\survarea} deg$^2$ roughly centred on the south Galactic pole. This is a region of sky with low brightness temperature, populated mostly by extragalactic sources, which we detect, catalogue and characterise in unprecedented detail at these frequencies.
The MWACS is the first systematic exploration of the end-to-end MWA system and the characteristics of the SKA site at these frequencies, and lays a base of understanding for the much larger and more comprehensive GaLactic and Extragalactic All-sky MWA (GLEAM) survey, which is currently underway and will survey the entire sky south of $\delta=+20^\circ$ between 72 and 230\,MHz (Wayth et al., in prep). In turn, this will lay a base for the massive continuum surveys that will take place at low frequencies with the SKA, from the same site.

This paper is laid out as follows.
\Sect~\ref{sec:obs_datared} describes the observations, and the calibration and imaging strategy used in data reduction.
We demonstrate how a hybrid strategy using conventional radio astronomy tools can generate high quality mosaics by taking advantage of the MWA's very good instantaneous $u,v$ coverage and near-coplanarity.
\Sect~\ref{sec:sourcefindflux} describes the source-finding and flux density calibration procedures, including correcting for the MWA primary beam and how we have dealt with resolved structures.
\Sect~\ref{sec:sourcecat} describes the properties of the source catalogue and discusses issues affecting reliability and sensitivity.
\Sect~\ref{sec:conclusions} contains a discussion and conclusion.

\section{Observations and data reduction}
\label{sec:obs_datared}
\begin{figure*}
\centering
    \begin{subfigure}[b]{0.3\textwidth}
                \label{fig:gamma_ants}
                \includegraphics[angle=-90,width=\textwidth]{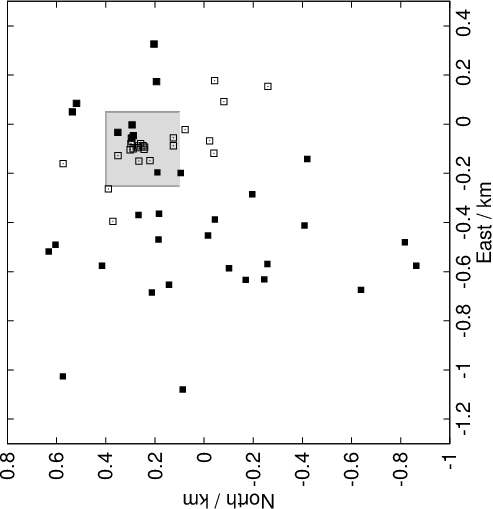}
    \end{subfigure}
    \begin{subfigure}[b]{0.3\textwidth}
                \label{fig:gamma_ants_zoom}
                \includegraphics[angle=-90,width=\textwidth]{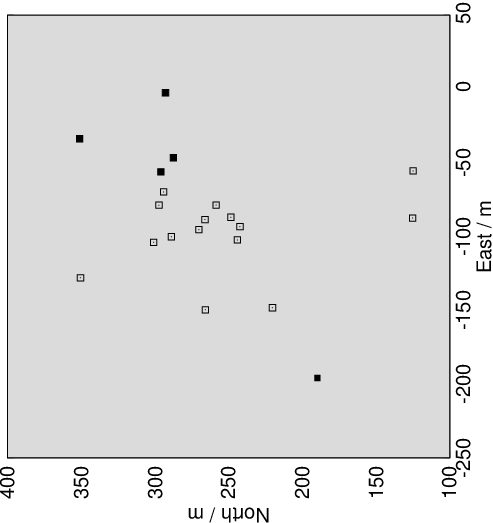}
    \end{subfigure}
    \begin{subfigure}[b]{0.3\textwidth}
                \label{fig:gamma_uv}
                \includegraphics[angle=-90,width=0.9\textwidth]{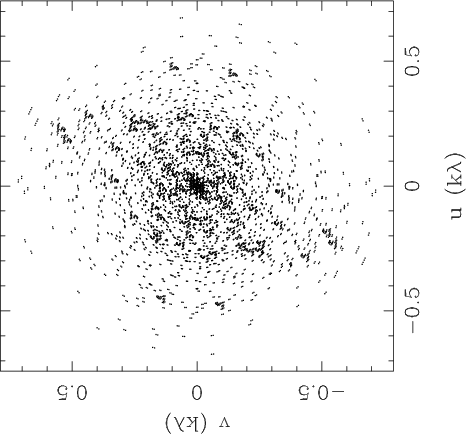}
    \end{subfigure}
    \caption{The left and middle panels show the antenna layout of \emph{beta} (empty squares) and \emph{gamma} (filled squares). The light gray shaded box on the left panel is enlarged in the middle panel, to more clearly display the central tiles. The monochromatic zenith-pointed snapshot combined $u,v$-coverage of the two MWA sub-arrays at 150\,MHz is shown on the right.}
    \label{fig:gamma}
\end{figure*}

As detailed by \citet{2013PASA...30....7T}, the MWA consists of 128 32-dipole antenna ``tiles'' distributed over an area approximately 3~km in diameter. Each tile observes two instrumental polarisations, ``X'' (16~dipoles oriented East-West) and ``Y'' (16~dipoles oriented North-South). The zenith field-of-view of a beamformed tile is $\approx30^\circ$ at 150\,MHz.
The signals from the tiles are collected by 16~in-field receiver units, each of which services eight tiles.
For engineering reasons, during the commissioning period only four receivers were active at any one time, hence the tiles and receivers were commissioned as an overlapping series of six 32-tile sub-arrays labelled \emph{alpha} through \emph{zeta}.
The sub-arrays were chosen to have good snapshot $u,v$-coverage within the various technical constraints in place during commissioning.
The data presented in this paper were recorded by sub-arrays \emph{beta} and \emph{gamma}.
The antenna layout and snapshot $u,v$-coverage of the combination of these two arrays are shown in \fig~\ref{fig:gamma}. The baselines are 8--1,530\,m in length, and their combined effective angular resolution at 180\,MHz is of order 3\arcmin (5\arcmin at 150\,MHz and 6\arcmin at 120\,MHz.). For simplicity, only the ``XX'' and ``YY'' polarisations are used in the following analysis; the cross-polarisation terms are discarded. The effect of ignoring these terms is constant throughout the night, as the instrument gains are very stable; any small loss in flux is later fixed during the flux calibration stage (Section~\ref{sect:abscorrect}).

A number of observation programs were undertaken during commissioning, one of which was night-time ``drift scans'' where the MWA tiles were pointed to a single Declination (Dec) along the meridian and data were collected as the sky drifted through the tile beams.
This form of observation has previously been shown to be an effective way to observe large fractions of the sky with good sensitivity for the MWA \citep{2013ApJ...771..105B} and many other instruments, past and present, that use fixed dipole arrays also use drift scans.
Using drift scans, uncertainty about the system calibration is minimised because the settings of all analogue components of the system are unchanged over the entire observation. The MWA has excellent stability of both the amplitude and phase of antenna gains using this mode, especially at night, when the ambient temperature changes slowly. We found that the standard deviation of primary-beam-corrected peak flux densities of typical bright unresolved ($>50$\,Jy) sources was only 1.5\% as they drifted through the zenith beam. Typical commissioning calibration scans show phase stability of better than than 1\arcdeg over the band.
Overall, we found the quality of drift scan data to be limited by the stability of the ionosphere, which generates slow astrometric changes (see \Sect~\ref{sec:astrometry}); in calm conditions a single calibration solution can be applied to all the data for an entire night, in the same fashion as \citeauthor{2013ApJ...771..105B}.

\begin{figure}
\centering
\includegraphics[width=0.5\textwidth]{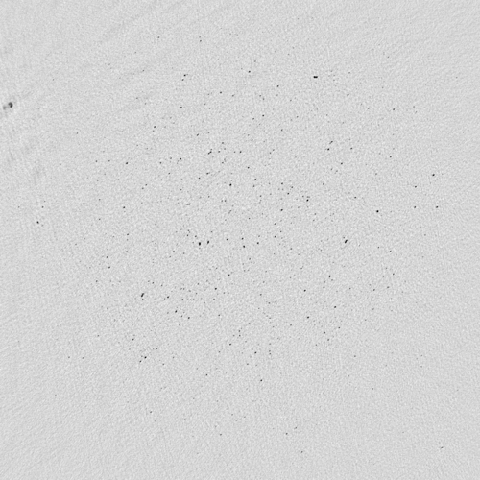} 
	\caption{An animation, at four frames per second, showing the central $30^\circ\times30^\circ$ of the Dec\,$-47$ 180\,MHz drift data created by combining the 180\,MHz visibilities of the nights of 2012~Oct~19 and 2012~Oct~31, and following the imaging procedure outlined in \sect~\ref{subsect:imaging}. In the usual convention, RA increases from right to left and Dec from bottom to top. The central Dec remains constant throughout at Dec$=-45^{\circ}35\arcmin35''$; the first frame is centred on RA$=21^{\mathrm{h}}26^{\mathrm{m}}57^{\mathrm{s}}$ and the last is centred on RA$=07^{\mathrm{h}}34^{\mathrm{m}}14^{\mathrm{s}}$. The colourscale is linear and runs from $-0.25$--1\,Jy\,beam$^{-1}$, but no correction has yet been made for the MWA primary beam. NB: Due to size limits, the animation is only visible in the published version of this article.}
    \label{fig:drift-movie}
\end{figure}

\begin{table*}
\centering
\caption{Summary of observations}
\label{tab:datasummary}
\begin{tabular}{lcccc}
\hline
Array & Central Dec & Date & Calibrator & Observing time / hours \\
\hline
\emph{beta} & $-26.7^\circ$ & 2012 Oct 18 & 3C444 & 9.8 \\
\emph{beta} & $-47.5^\circ$ & 2012 Oct 19 & Pictor~A & 11.1 \\
\emph{gamma} & $-26.7^\circ$ & 2012 Oct 30 & 3C444 & 11.0 \\
\emph{gamma} & $-47.5^\circ$ & 2012 Oct 31 & Pictor~A & 11.0 \\
\hline
\end{tabular}
\medskip\\
$^a$All observations used three frequency settings centred on 119.04, 149.76 and 180.48\,MHz with 30.72\,MHz bandwidth.\\
\end{table*}
%
%
The data presented in this paper are from drift scans taken at two Dec settings: $\delta=-26\fdg7$ (the zenith; hereafter referred to as Dec\,$-27$) and $\delta=-47\fdg5$ (hereafter referred to as Dec\,$-47$).
The observations are broken into a series of scans that cycle between three frequency settings with centre frequencies 119.04, 149.76 and 180.48\,MHz (henceforth given as 120, 150 and 180\,MHz); the scans are two minutes long and are separated by eight-second gaps.
After each scan, the frequency is changed and a new scan begins.
The bandwidth of the MWA is 30.72\,MHz, hence the total frequency range of our observations is 104 to 196\,MHz.
Observations at a particular frequency are thus separated into many two-minute scans that begin every six~minutes.
In October, the night-time drift field-of-view encompasses Right Ascension (RA) in the range $\approx20$h--$09$h; we restricted our analysis to the range where both \emph{beta} and \emph{gamma} data were available at identical local sidereal times: $21\mathrm{h}<\mathrm{RA}<8\mathrm{h}$.
The observations are summarised in \tab~\ref{tab:datasummary}, and in total comprised 3\,TB of unaveraged visibilities (reduced to 540\,GB after flagging and averaging; see \Sect~\ref{subsect:flagandcal}). \fig~\ref{fig:drift-movie} shows an animation of snapshots of one frequency produced by a typical drift scan.

Unlike \citet{2013ApJ...771..105B}, whose focus was the large-scale polarisation features of the low-frequency sky, these commissioning data are more suitable for measuring the flux densities of individual faint compact sources. \citeauthor{2013ApJ...771..105B} used the Real Time System \citep{2008ISTSP...2..707M,2010PASP..122.1353O} and a forward-modelling scheme to peel 250~sources from 2,400~square degrees of sky; given the instantaneous sensitivity of the commissioning sub-arrays, following such a procedure for our larger, more resolved sky, was infeasible. The following data reduction therefore uses standard astronomy tools such as Common Astronomy Software Applications (\textsc{casa}\footnote{http://casa.nrao.edu/}) and \textsc{miriad} \citep{1995ASPC...77..433S}.

\subsection{Flagging and calibration}\label{subsect:flagandcal}

The MWA band is comprised of 3072~10\,kHz `fine' channels, derived from 24~1.28\,MHz `coarse' channels.
The eight edge fine channels of each coarse channel suffer aliasing and are flagged in every observation.
The central fine channel of each coarse channel contains the (non-zero) DC component of the polyphase filterbank, so is also flagged.
Known misbehaving antennas were flagged, and the Orbcomm transmission frequency range 136--138.5\,MHz was completely excised.
Radio-frequency interference was excised using \textsc{aoflagger} \citep{2012A+A...539A..95O}; due to the radio-quiet nature of the observatory, less than three~per~cent of the data were affected and removed, mostly at the 136--138\,MHz range of the Orbcomm satellite downlink.
From a native frequency resolution of 10\,kHz, the data were averaged by a factor of four in frequency, appropriately downweighting any frequency bin containing a reduced number of fine channels. From the original time-sampling of 0.5\,s, \emph{gamma} data were averaged by a further factor of two in time, while \emph{beta} data were time-averaged by a factor of eight, the difference arising from the larger number of long baselines in \emph{gamma}.

A single calibration solution was determined for each frequency, for each night, based on the calibrator listed in Table~\ref{tab:datasummary}.

\subsubsection{Dec\,$-27$ phase calibration: 3C444}
The Dec\,$-27$ drift scans were phase-calibrated using 3C444: RA$=22^{\mathrm{h}}14^{\mathrm{m}}25.752^{\mathrm{s}}$; Dec$=-17^{\circ}01\arcmin36\farcs29$, which has a flux density $80$\,Jy and spectral index $-0.9$ at 160\,MHz \citep{1995AuJPh..48..143S}. 3C444 dominates the visibilities when observed near the meridian, and is unresolved at the resolution of the commissioning array. A two-minute observation produces solutions of S/N$\simeq5$, while integrating over the three two-minute snapshots in which 3C444 was closest to the meridian increases the S/N to $\simeq7$. As the sky has drifted by a total of 18\,minutes by the time data have been taken at all three frequencies, the instrumental response is different for each snapshot, so the expected flux of 3C444 was scaled by a primary beam model for each snapshot  (see \Sect~\ref{sect:abscorrect} for more details on the primary beam model). A least-squares fit over the full duration of each observation was performed on those visibilities with $u,v$-distance\,$>0.1$\,k$\lambda$ using \texttt{bandpass}, a \textsc{casa} routine, to produce a single complex gain solution for each tile, for each polarisation, for each frequency interval, which was then applied to the night's observations.

\subsubsection{Dec\,$-47$ phase calibration: Pictor~A}

The Dec\,$-47$ drift scans were calibrated using Pictor~A: RA$=05^{\mathrm{h}}19^{\mathrm{m}}49.7^{\mathrm{s}}$; Dec$=-45^{\circ}46\arcmin43\farcs70$. Pictor~A was used only as a phase calibrator; its flux density of $S\simeq400$\,Jy \citep{1995AuJPh..48..143S} over the MWA observing band dominates the visibilities such that a single two-minute observation is sufficient to obtain per-channel solutions with S/N$\gg20$. Since Pictor~A is resolved by our instrument, the Very Large Array (VLA) images at 1400 and 333\,MHz \citep{1997A+A...328...12P} were used to extrapolate a spatial and spectral model for each MWA observing frequency\footnote{While the VLA 74\,MHz image is closer to the MWA observing band, it possesses insufficient North-South resolution to accurately model the source structure, and therefore provide a model to properly phase-calibrate the MWA visibilities.}. The same procedure was followed as with 3C444: the model was Fourier-transformed and used to generate a calibration solution via \texttt{bandpass}, which was then applied to the night's observations.

After calibration, the \emph{beta} and \emph{gamma} data were concatenated. We note that the ionosphere was different between the two nights, but only the \emph{gamma} data posseses baselines of sufficient length to experience phase offsets significant compared to the width of the synthesised beam. Therefore it is possible to combine the \emph{beta} and \emph{gamma} visibility data for the same area of sky, but not to combine all of the \emph{gamma} data together, primary beam effects aside.

\subsection{Subtraction of bright sources in primary-beam sidelobes}

The first sidelobes of the MWA primary beam are estimated to have a response of approximately 3\% of that of the main beam, and the second around 0.1\% (see \Sect~\ref{sect:abscorrect} for more details on the primary beam model). When an extremely bright source such as Cygnus~A passes through such a sidelobe, it contributes signal to the visibilities. If not subtracted directly from the visibilities (``peeled'') or deconvolved, this contribution is visible in the image as striping from the sidelobes of the synthesised beam at the position of the source. In the absence of computing constraints, one could attempt to image the entire sky, such that all sources are deconvolved.
However, we found that bright sources sufficiently far ($>45^{\circ}$) from the phase centre were poorly deconvolved even in a snapshot image due to the array not being perfectly coplanar. At the time of analysis, techniques such as $w$-projection \citep{2008ISTSP...2..647C} were too computationally expensive for the $\sim$10,000~pixel-wide grid required to cover the sky. The sidelobes also introduce a large spectral variation to the image pixels, as they vary much more quickly with frequency than the main beam.

To reduce the effects of bright sources in the sidelobes of the primary beam, we phase-rotated the calibrated, concatenated snapshot datasets to the source positions and performed a shallow (threshold\,$=10$\,Jy) \textsc{clean} of a $1^{\circ}\times1^{\circ}$ around each source, with an additional Taylor term, allowing each pixel to have its own spectral index \citep{1994A+AS..108..585S}. The model pixels were then subtracted from the visibilities, which were then phase-rotated back to their original positions, ready for imaging.
The sources removed in this way were Pictor~A, the Crab Nebula, Hydra~A, Cygnus~A, PKS$2153-69$ and PKS$2356-61$. Only Pictor~A was present in the primary beam main lobe, as well as the sidelobe,
and we note that this strategy means we are then unable to recover its flux density in the final mosaic.

\subsection{Imaging strategy}\label{subsect:imaging}
At 150\,MHz, the full-width half maximum (FWHM) of the MWA primary beam at zenith is approximately $30^\circ$, corresponding to more than two hours of RA.
Sources therefore appear in many of the two~minute scans as the sky drifts through the beam.
Within any two-minute scan (hereafter a ``snapshot''), the hybrid \emph{beta}/\emph{gamma} array is sufficiently coplanar that snapshot images do not suffer from wide-field effects within the main primary beam lobe ($<15^\circ$ from the phase centre) and the image coordinates can be described by a slant-orthographic (generalised SIN) projection \citep{1999ASPC..180..383P,2002A+A...395.1077C}. The data were thus divided into snapshots with the phase centre defined as the local sidereal time in the middle of the snapshot.
The goal of data processing is to form an image of the entire RA range of the observation using all of the available data.

Given the drift scan observation strategy, standard synthesis imaging techniques are not suitable.
In a drift scan, instead of pointing the beam at adjacent locations on the sky, the beam stays constant and the sky moves, relatively.
The data are therefore best processed using a mosaicing methodology.
Typically the goal of a mosaic observation is to image objects larger than the primary beam of the telescope \citep[e.g.][]{1999ASPC..180..401H}.
Much of the mosaicing literature and software is therefore focused on recovering the large-scale structure in a synthesis image using an interferometer with very short baselines.
In the case of these data, the commissioning sub-array has only 12 baselines shorter than 20\,m and thus sensitive to scales $>5\arcdeg$, unlike the full MWA which fully samples up to scales of $12\arcdeg$ (at 180\,MHz). As we are also observing a part of the sky mostly devoid of objects of large angular scale, our focus is on combining the data to maximise sensitivity to compact structures.

To form the mosaic images, we perform the following steps, for each frequency band, for each polarisation:
\begin{itemize}
\item generate a output accumulation image for the data and primary beam in an equal area coordinate system with 0\farcm5 pixel resolution and over a sufficient region of the sky for all data;
\item for each snapshot:
\begin{itemize}
\item invert using multi-frequency synthesis over a field-of-view of $40^{\circ}\times40^\circ$, with a pixel resolution of 1\arcmin, weighting each $u,v$-cell equally (``uniform weighting'');
\item \textsc{clean} to a threshold of three times the typical snapshot RMS, which itself is typically $\simeq240$--80\,mJy\,beam$^{-1}$ from 120--180\,MHz; the synthesised beam sidelobes are only $\approx10\%$ of the peak response, and a gain factor of $0.1$ is used, thus avoiding \textsc{clean}ing the noise, and \textsc{clean} bias;
\item restore using an elliptical Gaussian approximation of the snapshot synthesised beam;
\item regrid and add the deconvolved image multiplied by the appropriate primary beam model into the accumulation data image;
\item likewise, regrid and add the square of the beam image into the accumulation beam image;
\end{itemize}
\item after all additions, divide the data image by the beam-squared image to form the final mosaic.
\end{itemize}

In essence, this follows equation (1) from \citet{1996A+AS..120..375S} which maximises the signal-to-noise ratio (S/N) in the output mosaic given the changing S/N over the field in each snapshot due to the primary beam.
We use the \textsc{miriad} for the imaging, deconvolution and image arithmetic, and use our own code to generate MWA primary beam models  (see \Sect~\ref{sect:abscorrect} for more details on the primary beam model). \textsc{miriad}'s understanding of advanced World Coordinate Systems (WCS) and ability to regrid between arbitrary coordinate frames is especially useful.
\begin{figure*}
	\centering
	\includegraphics[width=0.8\textwidth]{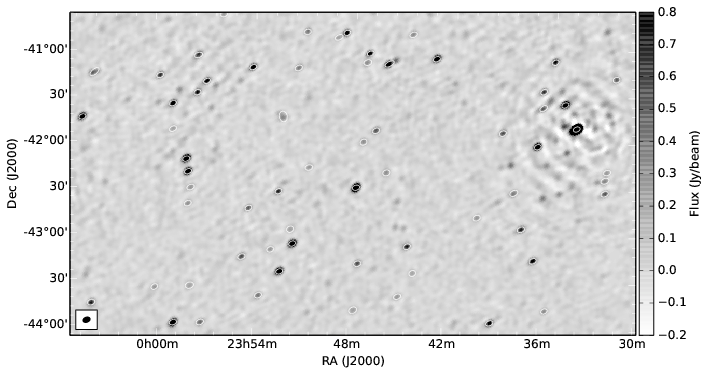}
	\caption{A randomly-chosen section of the 180-MHz Dec\,$-47$ pseudo-Stokes~I mosaic, representing about 4\% of the total MWACS survey area (in one of three frequency bands). The greyscale is linear and runs from $-0.2$ to $+0.8$\,Jy, and the estimated PSF at the centre of the image is shown as a boxed filled ellipse at the bottom-left of the image, of dimensions $4\farcm31\times2\farcm94$, position angle $-50\fdg4$. Some calibration errors are still in evidence, particularly around the bright source in the North-West. MWACS detections are shown as white ellipses of the same shapes as their fitted Gaussian parameters. The RMS of this image is 28\,mJy.}
    \label{fig:example-mosaic}
\end{figure*}

The process of adding snapshots improves the quality of the final image by reducing the thermal noise and improving the synthesised beam, revealing many fainter sources. These fainter sources have not been deconvolved in the snapshots, so their sidelobes remain in the mosaic with $\sim$1~hour of effective earth rotation synthesis (the time it takes for sources to move through the utilised area of the primary beam).
We estimate the ($5\sigma$) classical confusion limit \citep{1974ApJ...188..279C} to be approximately 5\,mJy\,beam$^{-1}$ given our effective synthesised beam size; this is consistent with the confusion measurement by \citet{2010A+A...522A..67B,2009A+A...500..965B} at 150\,MHz at a similar angular resolution and frequency.
While we therefore begin to approach the classical confusion limit, the dominant components of noise in our mosaics are sidelobe confusion and calibration errors. A randomly-chosen section of the 180-MHz Dec\,$-47$ mosaic is shown in \fig~\ref{fig:example-mosaic}, showing the well-behaved point spread function (PSF) and averaging down of calibration errors that results from the mosaicking process.

The zenith-pointed MWA primary beam is virtually identical for the XX and YY polarisations. We therefore averaged the XX and YY Dec\,$-27$ mosaics to form a (pseudo) Stokes-I image. For the Dec\,$-47$ mosaic, we found that the primary beam response of the XX and YY polarisations was sufficiently different that Dec-dependent corrections were required. Hence, the XX and YY mosaics for Dec\,$-47$ were not combined at this stage. Details of this process and absolute calibration are discussed in \sect~\ref{sect:abscorrect}.

\begin{table}
\centering
\caption{Gaussian parameters for corrected PSFs for the mosaics, at their original phase centres. The Dec\,$-47$ entries apply to both the XX and YY mosaics.}
\label{tab:psf}
\begin{tabular}{ccccc}
\hline 
Dec & $\nu$/MHz & $a$/' & $b$/' & $\theta$/$^{\circ}$\tabularnewline
\hline 
-27 & 119 & 6.17 & 4.30 & -63.8 \tabularnewline
-27 & 150 & 5.25 & 3.30 & -63.7 \tabularnewline
-27 & 180 & 4.41 & 2.79 & -63.6 \tabularnewline
-47 & 119 & 6.44 & 4.45 & -50.3 \tabularnewline
-47 & 150 & 5.08 & 3.47 & -50.4 \tabularnewline
-47 & 180 & 4.31 & 2.94 & -50.4 \tabularnewline
\hline 
\end{tabular}
\end{table}
\section{Source finding and flux-density estimation}
\label{sec:sourcefindflux}
The imaging process results in nine mosaics, consisting of three frequencies for the Dec\,$-27$ scan, and three frequencies $\times$ two polarisations for the Dec\,$-47$ scan. The RMS noise level of these mosaics is typically a factor of five lower than that of the individual snapshots, allowing much deeper source-finding; see \sect~\ref{sect:noise} for a discussion of the origins of noise in the mosaics, and their effect on the accuracy of our source flux densities.
\subsection{Sensitivity}
\begin{figure*}[t]
	\centering
	\begin{subfigure}{\textwidth}
                \centering
                \includegraphics[width=0.9\textwidth]{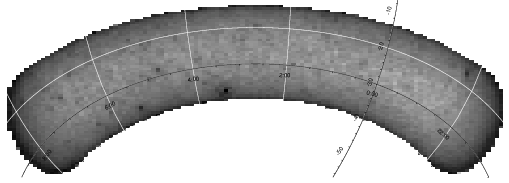}
                \label{fig:c102_rms}
	\end{subfigure}
	\begin{subfigure}{\textwidth}
                \centering
                \includegraphics[width=0.9\textwidth]{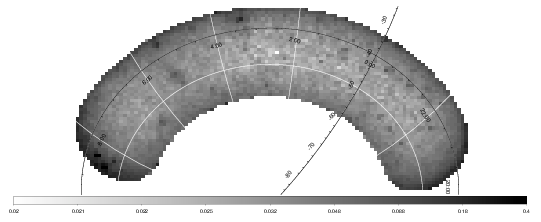}
                \label{fig:c103_rms}
	\end{subfigure}
	\caption{Log-scaled images of the root-mean-square (RMS) intensity measured across the mosaics for the highest-frequency Dec\,$-27$ (top) and Dec\,$-47$ (bottom) scan. The greyscale runs from 0.02 to 0.4~Jy\,beam$^{-1}$. The North--South axis is Dec in decimal degrees, and the East-West axis is RA in hours.}
    \label{fig:rmsmap}
\end{figure*}

Maps of the root-mean-square (RMS) value of the 180\,MHz mosaic pixels were generated using \textsc{Aegean} \citep{2012MNRAS.422.1812H} which calculates the RMS for blocks of pixels with dimensions of approximately 20$\times$20 beams\footnote{The RMS is calculated as the Inter-Quartile Range scaled by 1.34896. This is equivalent to the RMS for a Gaussian distribution but is more robust against high-flux pixels from bright sources \citep[c.f. e.g.][\sect~5.2.1]{1998AJ....115.1693C}.}.
These are reproduced for two representative mosaics in \fig~\ref{fig:rmsmap} (see also \fig~\ref{fig:rms_cdf}).
Two effects can clearly be seen in these figures: an increase in RMS at the edges of the mosaics, and isolated patches of locally high RMS.
The first is due to the effect of the primary beam and the second is due to the effect of sidelobe confusion and calibration errors around the brighter sources.
The latter is particularly severe near extremely bright, resolved structures which can most easily be seen close to the Galactic plane (at the Easternmost edge of the Dec\,$-47$ scan).
Fornax~A produces more calibration-error noise in the Dec\,$-27$ scan, and less in the better-calibrated Dec\,$-47$ scan; deconvolution errors from poorly reconstructing the largest scales of its structure are present in both.

\subsection{Source identification}\label{subsect:source-ident}

We wish to identify and characterise the morphology and flux density of all sources within the field which are bright enough to be discerned from the background noise. 
Having characterised the noise properties of the mosaics, we identify those pixels of the mosaics whose flux is greater than five times the local RMS, which itself is hereafter denoted $\sigma$.
\textsc{Blobcat} \citep{2012MNRAS.425..979H} was used to identify `blobs': islands of pixels containing a $5\sigma$ peak, expanded to include all contiguous pixels brighter than $3.5\sigma$, while accounting for various biases.

This was repeated for all three (Dec\,$-27$, Dec\,$-47$ XX and Dec\,$-47$ YY) of the highest frequency (180\,MHz) mosaics.
These were used exclusively for source-finding since the highest resolution allowed the greatest separation of neighbouring sources into discrete blobs.
The objects detected in the three maps were combined into a single list (i.e. from this point on no regard was taken as to which map the source was detected in).

\subsection{Characterising the point spread function}
For an interferometer coplanar to the plane perpendicular to the zenith, observing at zero hour angle, the synthesised beam becomes elongated N-S relative to the zenith, as $\csc\zeta$, where $\zeta$ is the zenith angle. Regridding from a slant orthographic projection to a zenithal equal area projection (using standard \textsc{miriad} WCS tools) conserves the synthesised beam volume, ignoring this effect, such that a correction factor of $\csc\zeta$ must be applied to rescale the N-S extent of the synthesised beam. We apply this immediately, before making further measurements.

While the synthesised beam shape for an individual snapshot is known very accurately (short-timescale ionospheric effects aside), determination of the effective synthesised beam in the mosaics is a much more difficult problem.
The mosaic contains approximately an hour (depending on frequency) of effective earth rotation synthesis which significantly circularises the PSF. However, refraction from the ionosphere acts differently in each snapshot and enlarges the PSF as snapshots are combined to form each mosaic.

\begin{figure}
	\centering
	\includegraphics[angle=-90,width=0.5\textwidth]{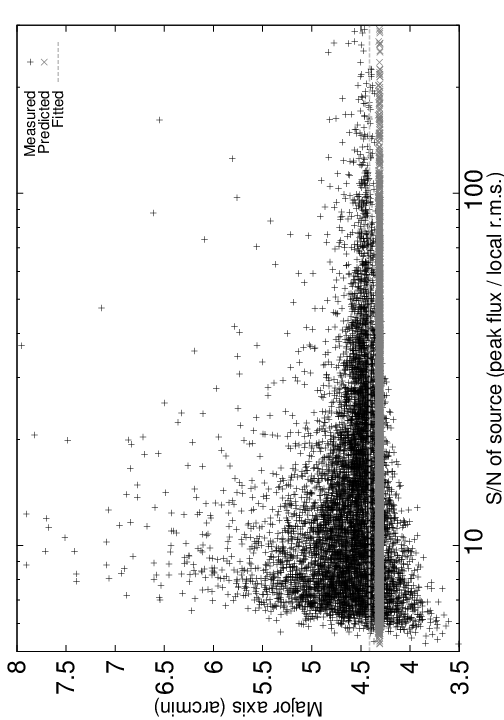}
	\caption{An example of the analysis used to find the correct PSF for the mosaicked images: in this case, we examine the major axis of sources detected in the zenith scan at 180\,MHz. The black crosses show the measured major axis of each source against its signal-to-noise ratio (S/N); the grey stars show the expected major axis if the zenith-angle-dependent projection effect were the only source of change in the synthesised beam; the light grey dashed line shows a S/N-weighted horizontal fit to the major axis measurements. The ratio of this fit to the predicted major axis gives the correction factor by which the major axis of the PSF must be increased to match the data: see \tab~\ref{tab:psf} for a list of the corrected PSFs for each mosaic.}
    \label{fig:SNR-a_funnel_plot}
\end{figure}

To measure this effect, we fitted 2D Gaussians to all detected sources, and plotted the distribution of ellipse values against various parameters (such as RA, Dec) and checked for correlations: none were seen. Plotting against S/N, as in \fig~\ref{fig:SNR-a_funnel_plot}, shows that the ellipse sizes of unresolved sources form a line, with the resolved sources having larger values. For each Gaussian parameter, a line was fit to the unresolved sources, and compared with the predicted value in order to calculate the correction needed to increase the size of the PSF to match the data.

\begin{figure}
	\centering
	\includegraphics[width=0.5\textwidth]{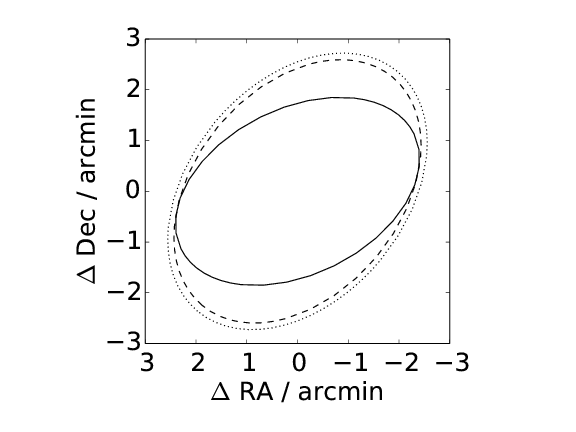}
	\caption{An exaggerated example of the corrections made to the PSF: the solid black ellipse shows a 150\,MHz synthesised beam at the zenith; the dashed line shows the synthesised beam for a pointing due South, at a zenith angle of $45^{\circ}$ (15$^\circ$ further than the maximum zenith angle of MWACS), resulting in a stretch in Dec; the dotted line shows an increase in the size of the PSF by a further 5\%; the real magnitude of the correction is closer to 2\%, and accounts for the effects of ionospheric smearing and image-based mosaicking.}
    \label{fig:beam_stretch}
\end{figure}

Henceforth, all source-finding uses the corrected PSFs, shown in \tab~\ref{tab:psf}. Note that the PSFs are identical for the XX and YY polarisation Dec\,$-47$ mosaics, as all of these effects are polarisation-independent. A further projection factor of $\csc\zeta$ is applied when measuring sources in the mosaics.
\fig~\ref{fig:beam_stretch} shows an exaggerated illustration of the effect of these changes. After these corrections, it is possible to recover the flux densities of extended sources, which would otherwise be over- or under-estimated depending on the difference between the local PSF and average PSF.\footnote{We note some similarity in this process to typical optical image processing, in which the PSF is determined by measurements of unresolved stars.}

\subsection{Characterisation of discrete sources}\label{subsect:sourcechar}
We now aim to determine flux density and morphology as accurately as possible given the three (Dec\,$-27$), six (Dec\,$-47$) or nine (overlapping region) measurements that we have for each source.
On visual inspection it is clear that the vast majority of our objects can be characterised by a single elliptical Gaussian similar in shape to the synthesised beam.
This reflects the fact that the vast majority of our sources are unresolved, as expected.

In order that they be fit with the minimum number of free parameters possible, we begin by modelling each of our sources as a single elliptical Gaussian, only attempting to characterise more complex morphology when this assumption is determined to be invalid (see \Sect~\ref{sect:goodbadsplit}).
In another refinement to avoid the effects of confusion due to nearby sources, the fits were performed only over those pixels which lie within the FWHM of the \textit{a priori} fit (with an extra margin of 10\%).
Levensburg-Macquart least-squares fits were made to these pixels, with free parameters: amplitude; major axis, $a$; minor axis, $b$; position angle, $\theta$; RA offset, $\Delta_\mathrm{RA}$; and Dec offset, $\Delta_\mathrm{Dec}$.
The initial parameters were the synthesised beam (with frequency and position dependence calculated as described in \sect~\ref{subsect:imaging}) with an amplitude of the peak pixel in the region of the fit, centred on the flux weighted centroid of the object.

For error calculation purposes the RMS in the region of the blob in each map was determined by taking the mean of the RMS map over an array of $24\times24$ pixels centred on the centroid of the blob.
This provides a modest amount of smoothing of the RMS shown in \fig~\ref{fig:rmsmap} (where the RMS is determined over blocks a factor of $\sim5\times5$ larger).

\subsubsection{Separating single-component and multi-component sources}
\label{sect:goodbadsplit}
Next we determine which of the blobs are indeed well-characterised by a single elliptical Gaussian and which require further characterisation.
Our criterion is that any source whose centroid lies within the half-maximum of the \textit{a priori} elliptical Gaussian is ``single component''. This simple criterion is effective due to our selection of the \textit{a priori} centroid on the basis of the amplitude-weighted centroid of the blob.

When there is no peak (i.e. positive curvature in both spatial dimensions) within our stringent pixel range, the centroid of the fit is forced outside the \textit{a priori} ellipse.
If there is a peak, but it is not coincident with the \textit{a priori} centroid, this is indicative of signal within the blob which is not associated with the fitted Gaussian.

Using this method, {\nunresolved} ($\approx91$\,\%) of our {\nsrc} sources are classified as single-component.

\subsection{Flux-density calibration}
\label{sect:abscorrect}
The southern sky at low frequencies is poorly surveyed and flux densities derived from extrapolation of various radio catalogues to the MWA frequency range disagree at the 10--20\% level. Work is under way in the community to develop a unified radio flux density scale from MHz--GHz frequencies\footnote{http://mwsky.ncra.tifr.res.in/mwsky/upload\_talk/11-Dec\_3B\_Rick\_1337\_y.pdf}, and GLEAM will expand this scale over the entire Southern sky. At the time of writing, the best course of action was to bootstrap from the fairly well-understood Northern sky to the South. Conveniently these declinations pass close to the zenith for the MWA, allowing the use of a simple model of the MWA primary beam.

The MWA primary beam has previously been approximated by an analytic model incorporating a dipole over groundscreen and geometric array factor with good results \citep[e.g.][]{2012ApJ...755...47W,2013MNRAS.436.1286M,2014MNRAS.438..352B}, and we adopt the same approach in this paper.
\citet{2013ApJ...771..105B} find that the zenith primary beam model is incorrect at the $\sim$2\% level at higher frequencies, and we expect the model to be less correct further from zenith as the simple dipole approximation becomes less valid.
Due to the uncertainty in the primary beam model, especially away from the zenith, it is impossible to correctly perform the absolute flux density calibration in a single step.
In addition, phase errors from imperfect phase calibration will reduce the recovered peak flux of radio sources, necessitating a small, position-independent correction.
We note that understanding and improving the primary beam model is an ongoing activity within the MWA collaboration.
Recent work \citep{2014Sutinjo} to incorporate mutual coupling effects into the model has improved our understanding of the response of the MWA beam.
However the work is still ongoing and for the purposes of this paper our empirical method to bootstrap the flux density scale, described below, was sufficient.
\begin{figure}
\centering
		\includegraphics[width=0.9\columnwidth,clip=]{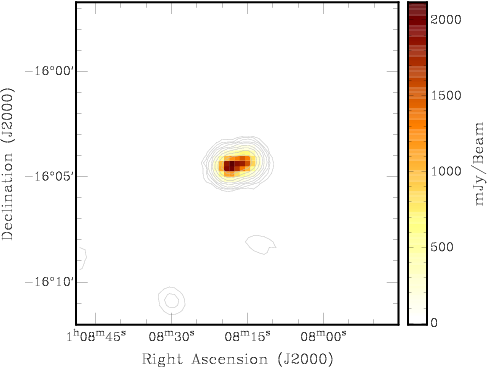}
		\caption{A postage stamp image of 3C32 extracted from the NVSS survey. Peak flux density is 2.11\,Jy\,beam$^{-1}$. Contours begin at 0.2\% peak and have a common ratio of 2.}
		\label{fig:3c32_nvss}
\end{figure}
\begin{figure}
\centering
		\includegraphics[width=1.1\columnwidth]{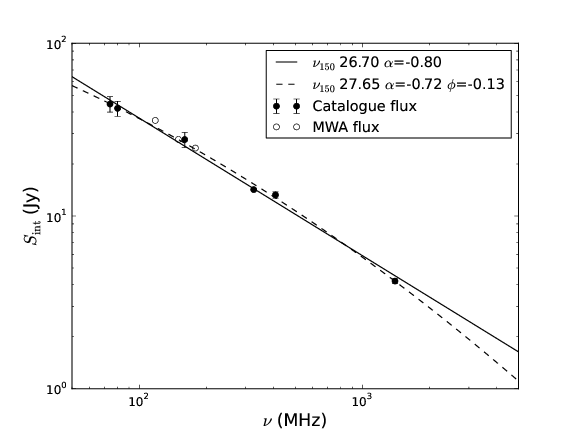}
		\caption{3C32 flux densities from the VLSS, Culgoora, MRC, and NVSS surveys shown with uncorrected MWA fluxes. Two fits to the non-MWA points are shown: the solid line is a power-law (i.e. a straight-line fit in log-log space); the dashed line fits a parabola in log-log space. MWA flux densities were corrected to lie on the power-law fit. }
		\label{fig:3c32_spectrum}
\end{figure}

In order to provide an absolute flux density scale, a list of sources was drawn up which appear in our catalogue as well as all of the Culgoora \citep[80, 160\,MHz][]{1995AuJPh..48..143S}, VLSS \citep[74\,MHz][]{2007AJ....134.1245C}, WISH \citep[325\,MHz][]{2002A&A...394...59D}, MRC \citep[408\,MHz][]{1981MNRAS.194..693L} and NVSS \citep[1.4\,GHz][]{1998AJ....115.1693C} catalogues.
A good choice of flux calibrator might be our northern phase calibrator, 3C444; however, there are two reasons to avoid using this source. 
Firstly, it is towards the edge of our RA range, and is only sampled by a few snapshots.
Secondly, it is resolved into two components by NVSS, so measurements by different instruments with different resolutions may give conflicting results.

\begin{figure*} 
	\centering
	\rotatebox{90}{
	        \begin{minipage}{\textheight}
			\begin{subfigure}[b]{0.3\textwidth}
				\label{fig:3c32_93}
				\includegraphics[width=1.0\textwidth]{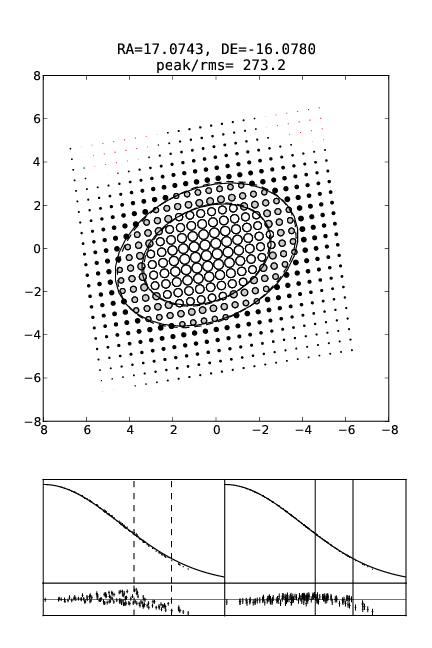}
			\end{subfigure}
			\begin{subfigure}[b]{0.3\textwidth}
				\label{fig:3c32_117}
				\includegraphics[width=1.0\textwidth]{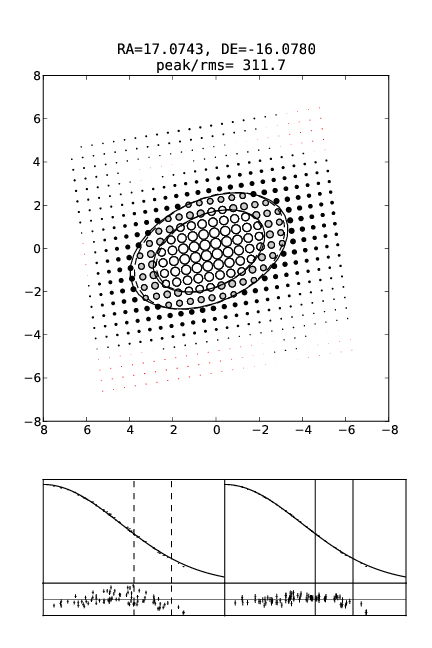}
			\end{subfigure}
			\begin{subfigure}[b]{0.3\textwidth}
				\label{fig:3c32_141}
				\includegraphics[width=1.0\textwidth]{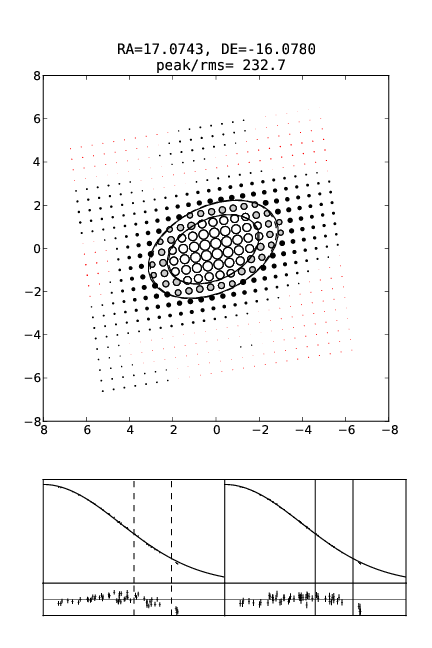}
			\end{subfigure}
				\caption{\label{fig:3c32_pixmap} Least-squares fits to pixel values for the source 3C32 at 119, 150 and 180\,MHz.
				RA and Dec are in degrees; axes of top figure are arcminutes.
				Circles represent image pixels with circle diameter proportion to brightness.
				Unfilled pixels are those with at least 50\% of the brightness of the \textsc{Blobcat} peak.
				Grey are those with at least 25\% of the \textsc{Blobcat} peak.
				Red are those with negative values.
				Solid ellipses show half-power-beam-width and $\sqrt{2}\times$half-power-beam-width of the fitted elliptical Gaussian.
				Dashed ellipses show the synthesised beam.
				Bottom-left plots and residuals show fit of pixels with $a$, $b$ and $\theta$ constrained to the synthesised beam parameters.
				Bottom-right plots and residuals show fit of pixels with $a$, $b$ and $\theta$ as free parameters.
				Vertical lines on bottom plots correspond to the ellipses on the upper plot.
				The fit ellipses are barely discernible from the synthesised beam ellipses, indicating that the sources is unresolved.
				}
		\end{minipage}
		}
\end{figure*}

The second brightest candidate is 3C32. 
This source as it appears in NVSS (see \fig~\ref{fig:3c32_nvss}) would be unresolved in our data; its flux density is well-modelled by a simple power law (\fig~\ref{fig:3c32_spectrum}) and shows no evidence of variability or being resolved.
In particular, the VLSS 74\,MHz and Culgoora 80\,MHz flux densities both lie on the power-law fit within the errors, despite their differing resolutions and epochs of observation. We performed an image-plane Gaussian fit (\fig~\ref{fig:3c32_pixmap}) to measure the flux density of the source and confirmed it as unresolved.
We therefore scaled the integrated flux densities of each of the Dec\,$-27$ maps appropriately to fit the least-squares fit to the catalogue fluxes (see \tab~\ref{tab:abscorrect}).
Since the zenith analytic primary beam model had already been divided out in creating the Dec\,$-27$ mosaic, the relative flux densities of sources were already correct and only this single scaling factor should be required to set the absolute flux density scale across the whole mosaic, assuming that the zenith primary beam model is correct.
\begin{table}
\centering
\caption{Scaling factors applied to the measured flux densities. The Dec\,$-27$ corrections make the fluxes consistent with the absolute flux scale determined for 3C32. The Dec\,$-47$ corrections make the flux densitites in the two fields consistent with each other.}
\label{tab:abscorrect}
\begin{tabular}{lcc}
\hline
$\nu$/MHz & Dec\,$-27$ Correction & Dec\,$-47$ Correction \\
\hline
119 & 0.904 & 0.600 \\ 
150 & 0.963 & 0.769 \\ 
180 & 0.939 & 0.818 \\ 
\hline
\end{tabular}
\end{table}

\begin{figure}
\centering
		\includegraphics[width=\columnwidth]{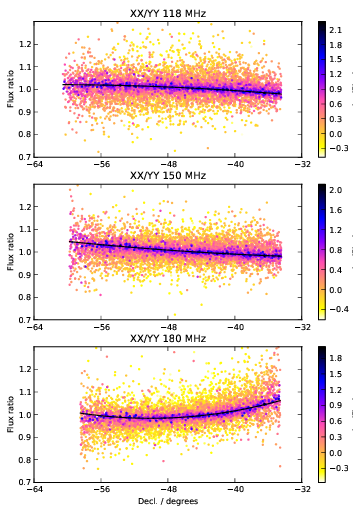}
		\caption{Ratio of XX to YY flux density for all sources detected in both maps plotted against Dec for all frequencies (low to high). The colour axis is the log$_{10}$ of the source flux density in Jy (as measured in the YY scan before amplitude calibration).}
		\label{fig:xx_yy}
\end{figure}

The flux densities measured in the XX and YY mosaics of the Dec\,$-47$ field disagree at the $\sim$5\% level due to unmodelled primary beam effects.
A second-order polynomial fit (weighted by the two flux densities added in quadrature) was made to the Dec-dependent discrepancy between XX and YY as shown in Figure~\ref{fig:xx_yy}.
The XX mosaic was then scaled to match the YY mosaic.

The Dec\,$-47$ mosaic was corrected to be consistent with the Dec\,$-27$ mosaic by using the $\approx600$ unresolved sources which lay in the overlapping region.
The mean of the flux density ratio between the two maps of all of these sources was taken to be the additional correction factor to the Dec\,$-47$ mosaic to produce a corrected absolute flux density calibration scale consistent with that used for Dec\,$-27$ (\Tab~\ref{tab:abscorrect}). After this correction, the XX and YY mosaics of the Dec\,$-47$ field were combined, weighted by their respective primary beam responses (which differ by around 5\%), to form a pseudo-Stokes-I mosaic, on which source-finding can be performed.

\subsection{Combining the mosaics}
Many MWACS sources are detected both in the Dec\,$-27$ and Dec\,$-47$ mosaics.  
Combining the information from both mosaics is likely to give superior results for a number of reasons.
This would recover some of the signal-to-noise in the overlapping region of the two mosaics (at Dec$\simeq-36^\circ$) which is otherwise attenuated by the primary beam, especially at the highest frequency.
The differing primary beam between the two scans also means that the signal from sources in the distant sidelobes of the primary beam will be somewhat different, so combining the mosaics would reduce this noise.

Unfortunately, the very different state of the ionosphere between the two \emph{gamma} drift scans makes a simple image-plane combination difficult; without extensive modelling of the ionosphere, there would be large position- and frequency-dependent smearing of the sources. This is evident from \Fig~\ref{fig:astrometry}; the astrometry errors are time-dependent (seen as a change in RA) and differ night-to-night.
However, an improved flux-density estimate can be determined for each source in the overlapping region by taking the mean of the two peak flux densities, weighted by the S/N of the fit in each of the maps (as defined in \sect~\ref{sect:errors}).

Though the size of the overlapping region varies dramatically with frequency, for simplicity and transparency the map flux densities were only combined in the region well-sampled by both the Dec\,$-27$ and Dec\,$-47$ scans at the highest frequency: a Dec range of $-38\fdg20 < \delta < -35\fdg25$. Furthermore, the positions and morphologies of the sources in the overlap region were taken as those measured in the Dec\,$-47$ mosaic, as this scan has a somewhat more compact synthesised beam and marginally better phase calibration (see \sect~\ref{sect:noise}).

\subsection{Fitting 180\,MHz flux densities}

The majority of the MWACS source-flux densities are well-characterised by a simple power-law spectral index; as described in \sect~\ref{subsect:source-ident}, we perform source detection at the highest resolution available, at 180\,MHz, so we use the fitted integrated source flux densities at each frequency to measure the spectral indices. Alongside this, we report the 180\,MHz flux density for each source.

We performed weighted least-squares fits to the flux densities at the three frequencies, propagating the confusion and fitting errors described in \sect~\ref{sect:errors}.

\subsection{Estimation of errors}
\label{sect:errors}

\subsubsection{Noise-like errors}\label{sect:noise-like-errors}
\citet{1997PASP..109..166C} addresses the problem of determining the error on the parameters of a least-squares fit of an elliptical Gaussian $G(x_k, y_k)$ to a set of pixel values $a_k$.
Here, we adopt the same notation where $A$, $\theta_M$, $\theta_m$, $\phi$ $x_0$ and $y_0$ are the amplitude, FWHM major and minor axes, position angle and centroid offsets, respectively, of a Gaussian fit over $a_k$ pixel amplitudes, each having the same Gaussian error distribution of RMS $\mu$.
In our case, the noise, whether it be thermal or confusion noise, will be correlated on the scale of the synthesised beam.
For the case of unresolved or weakly resolved sources, where the correlation of the noise matches the apparent source size, the error on $A^2$, $\mu^2(A)$ is $\mu^2$, i.e. the error on the flux density of the source due to noise is simply the local RMS.
The errors on $A$, $\theta_M$ and $\theta_m$ are therefore:

\begin{equation}
	\label{eqn:err_fit}
	\frac{\mu^2(A)}{A^2} = \frac{\mu^2(\theta_M)}{\theta_M^2} =  \frac{\mu^2(\theta_m)}{\theta_m^2} = \frac{\mu^2}{A^2} .
\end{equation}
By analogy to \citet{1997PASP..109..166C} \eqn~21, 
\begin{equation}
	\mu^2(\phi) = 2\left(\frac{\theta_M\theta_m}{\theta_M^2-\theta_m^2}\right) \frac{\mu^2(A)}{A^2} .
\end{equation}
Where $\mu(\phi) > 90^{\circ}$, $\mu(\phi)$ is set to null.

Finally, the errors on RA ($\alpha$) and Dec ($\delta$) due to noise-like errors are given by

\begin{eqnarray}
	\mu^2(\alpha) &=& \mu^2(\theta_M)\sin^2(\phi) + \mu^2(\theta_m)\cos^2(\phi) ,\\
	\mu^2(\delta) &=& \mu^2(\theta_M)\cos^2(\phi) + \mu^2(\theta_m)\sin^2(\phi) .
\end{eqnarray}

\subsubsection{Calibration errors}
In the bright source regime, the error on the flux density is dominated not by noise but by calibration errors.
This is added in quadrature to the other sources of error as a fractional error of 4\% on the integrated flux density. This was determined by measuring the local increase in RMS flux density around bright sources of known flux density, where calibration errors are the dominant source of noise, and dividing it by the flux density of each bright source.

\subsection{Remaining systematic errors}
\subsubsection{Error on the absolute calibration}
\label{sect:abserr}

\begin{figure*}[p]
	\centering
	\includegraphics[width=0.8\textwidth]{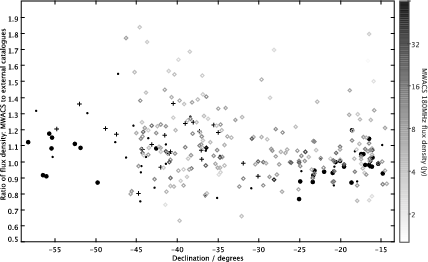}
	\caption{Plotted against Dec, the ratio of 180\,MHz flux density as predicted from catalogued values to that determined in our survey, for cross-matched unresolved sources. Unfilled diamonds represent 210 unresolved Culgoora sources with the 160\,MHz flux density scaled to a 180\,MHz flux density using the MWACS spectral index. Greyscale is the MWACS source flux density in Jy. Other points are fits to sources in the MRC and MS4 samples (as described in \sect~\ref{sect:abserr}). Dots represent the 32~sources fit best by a curved spectrum. Crosses show 21~sources without curved spectra whose 180\,MHz flux was extrapolated. Circles show the 32~sources classified as neither curved nor extrapolated.} 
    \label{fig:bh}
\end{figure*}
\begin{figure*}[p]
	\centering
	\begin{subfigure}[b]{0.49\textwidth}
                \centering
                \includegraphics[width=\textwidth]{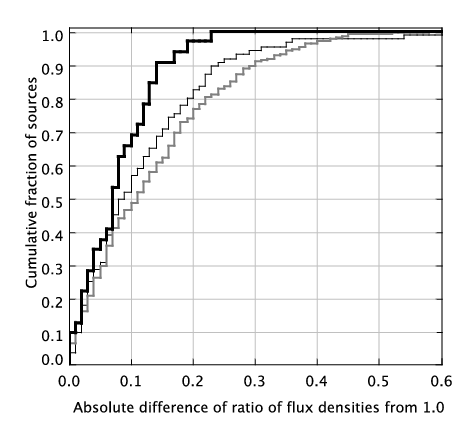}
	\end{subfigure}
	\begin{subfigure}[b]{0.49\textwidth}
                \centering
                \includegraphics[width=\textwidth]{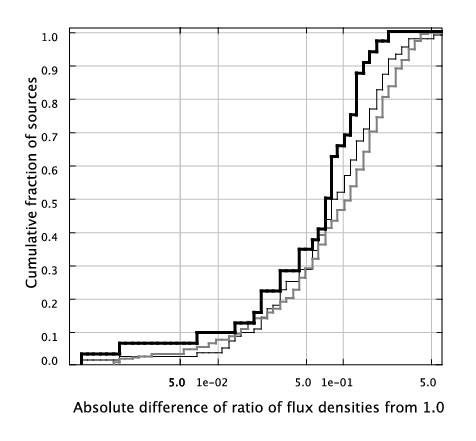}
        \end{subfigure}
\caption{The cumulative histograms of the difference of the flux density ratio from unity, where the ratio is the ratio of the flux densities of unresolved sources found in MWACS to the flux densities as measured by other catalogues (see \Sect~\ref{sect:abserr}). The grey line shows Culgoora, the thin black line shows all MS4 and MRC4 sources, and the thick black line shows only those MS4 and MRC4 sources which do not have curved spectra and are not pure extrapolations from 408\,MHz downward. The left panel shows the histogram in linear intervals, while the right panel shows the histogram with a log scale.\label{fig:flux_ratio_histogram}}
\end{figure*}

To test the consistency of the absolute flux density calibration across the field and across different sources, we first crossmatched our catalogue with the only other catalogue which lies within our frequency range: the Culgoora-3 list of radio source measurements at 160\,MHz, at a resolution of 1\farcm85.
These flux densities were scaled to 180\,MHz using the MWACS spectral index and the scaled flux densities compared with the MWACS flux densities, shown in \fig~\ref{fig:bh}.
While the scatter in the ratio of flux densities is large, and considerably larger than the estimated error on the Culgoora fluxes \citep[][\tabs~1\&2]{1977AuJPA..43....1S}, there is no evidence of a systematic offset with respect to Declination, which would have indicated uncorrected primary beam errors.

In addition, we identified all 41 sources which appeared in all the five catalogues used for absolute calibration (\sect~\ref{sect:abscorrect}) and which had an MRC (408\,MHz) flux density greater than 4\,Jy  (hereafter the MRC4 sample).
Since these catalogues only cover a restricted Dec range near the top of our field, these sources were supplemented by 73 sources from the ``MS4'' sample, defined by \citet{2006AJ....131..100B} as those sources in the MRC catalogue with a flux density greater than 4\,Jy.\footnote{In \Sect 4 of \citet{2006AJ....131..100B}, the authors present extrapolated 178\,MHz flux densities; we do not make use of these data.}

From the resulting sample we excluded all sources which were resolved in our catalogue -- defined as those with a fitted ellipse $>5$\% larger than synthesised beam ellipse (19/73 MRC4 sources and 9/41 MS4 sources were excluded).
Two fits were made to the catalogued flux densities: one a simple power law (i.e. a straight-line fit in log-log space); another incorporating curvature (a parabolic fit in log-log space).
Those sources where the 180\,MHz flux density predicted by the two fits were discrepant by more than 10\% were classified as ``curved''.
For the MS4 sources we used the multi-frequency flux density measurements compiled by \citet{2006AJ....131..100B}.
Many of these sources had no listed flux density below 300\,MHz (i.e. the 180\,MHz flux density is extrapolated from flux density measurements at 365 or 408\,MHz and above).
These sources were classified as ``extrapolated''.
The resulting flux density ratios are also shown in \fig~\ref{fig:bh}, which overall illustrates the challenge of defining a low-frequency flux scale in the Southern Hemisphere. There are few sources with well-characterised spectra at this frequency, and there are inconsistencies between these sources.

\fig~\ref{fig:flux_ratio_histogram} replots these data as cumulative histograms of the difference of the flux density ratio from unity. Three subsets of (unresolved) sources are shown: Culgoora, MS4 \& MRC4, and MS4 \& MRC4 sources which do not have curved spectra and are not pure extrapolations from 408\,MHz downward. Using this latter, most high-quality subset, we see that 68\% of sources lie within 10\% of unity. We therefore add an absolute flux calibration error of 10\% in quadrature to the final flux density error after fitting.
 
\subsubsection{Primary-beam errors}
For a mosaicked drift scan, an error in the primary-beam model would be expected to manifest itself as a Dec-dependent error in flux density.
Overall there is little evidence of a Dec-dependent error in \fig~\ref{fig:bh}. 

\begin{figure}
	\centering
	\includegraphics[width=\columnwidth]{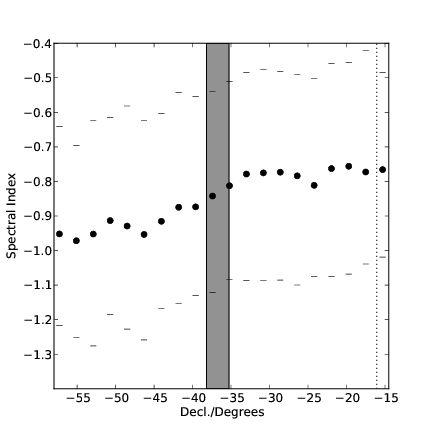}
	\caption{Quartiles of spectral index (filled circles indicate median, bars indicate upper and lower quartiles) of sources in each of 20 equal-width bins covering the full range of Dec. The shaded region indicates the declination range of the overlap region; the dotted vertical line indicates the declination of 3C32.} 
    \label{fig:alpha_de}
\end{figure}

Any error in the primary beam model would very likely have strong frequency-dependence, such that it may systematically affect the spectral index.
This can be tested by binning the sources by spectral index and looking for systematic changes with Dec.
The results of this analysis are shown in \fig~\ref{fig:alpha_de}.

There are clearly systematic changes in the median spectral index with Dec, though there is no obvious functional form which could be modelled to take account of this.
However the change in spectral index across the band corresponds to an error across the frequency range of less than 10\%. 
An error of 0.1 is added in quadrature to the to the error on the fitted spectral index to take account of the systematic shift in spectral index seen here.

\subsubsection{Astrometry errors}\label{sec:astrometry}
\begin{figure}[h!]
        \includegraphics[width=0.5\textwidth]{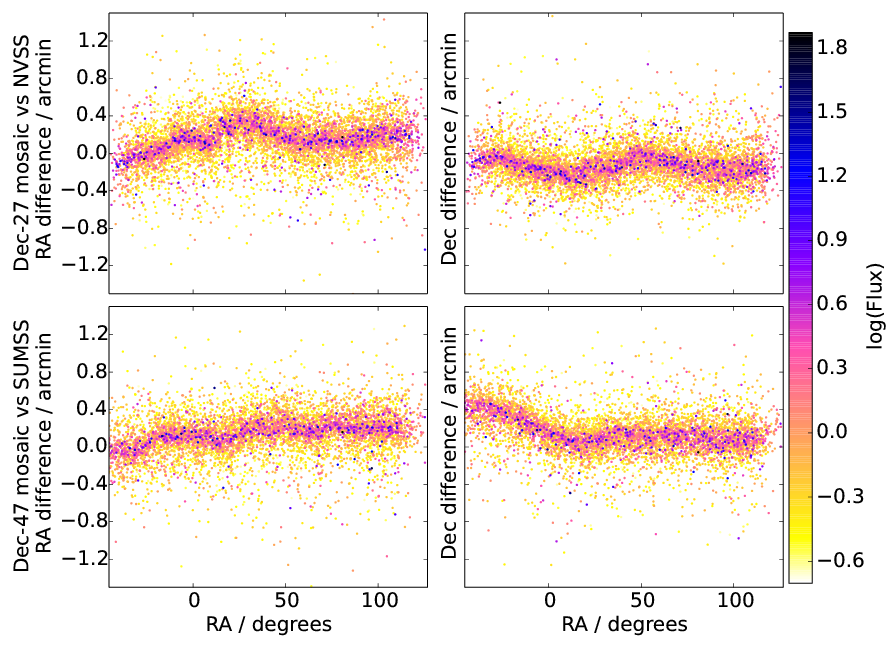}
	\caption{Astrometric offsets (left: RA; right: Dec) for the 180\,MHz mosaics, shown against RA (effectively, time of night). Top: Dec\,$-27$ mosaic, cross-matched against NVSS. Bottom: Dec\,$-47$ mosaic, cross-matched against SUMSS. The colour axis is the log$_{10}$ of the source flux density in Jy (as measured in the 180\,MHz mosaic before amplitude calibration).}
    \label{fig:astrometry}
\end{figure}

\Fig~\ref{fig:astrometry} shows the astrometric error for all unresolved sources cross-matched between our catalogue and the NVSS catalogue for the Dec\,$-27$ mosaic, and the SUMSS catalogue for the Dec\,$-47$ mosaic.
The astrometric errors in RA and Dec are not entirely correlated, vary with RA (time, for a drift scan), and are likely to be functions of ionospheric activity and viewing angle through the ionosphere (i.e. zenith angle). As we do not attempte to solve for this systematic error, we measure the RMS without weighting each measurement by the flux density of the sources, which leads to a conservative astrometric RMS estimate of 0\farcm3.
This value is added in quadrature to the errors on the RA and Dec derived in \sect~\ref{sect:noise-like-errors}.
%
\subsection{Resolved sources}\label{sect:multi-comps}
While the majority of sources in the catalogue are unresolved at the resolution of the commissioning array, and are therefore described well by a single flux density, some objects are more complex and must be examined more closely to determine an integrated flux density. These are automatically detected as described in \sect~\ref{sect:goodbadsplit}.

For each of these sources, a $100\times100$ pixel subsection of the mosaic was regridded to a local SIN projection, and the appropriate point spread function determined in \sect~\ref{subsect:imaging} was added to the header. The source-finding software \textsc{Aegean} was then run on the sub-image, detecting and fitting multiple components simultaneously.
The results of this fitting at the highest frequency only were used to produce the integrated flux density measurement at 180\,MHz.

Due to the varying resolution over the bandwidth of the array, cross-matching from the 180\,MHz-detected sources to the other two frequencies is not always straight-forward when sources are extended, or close to each other.
\begin{itemize}
\item{For some sources, particularly those which are only slightly extended and not confused with other nearby sources, the components match easily, and integrated flux densities are calculated for each freqency.}
\item{For confused sources, where different numbers of components are detected at different frequencies, an integrated flux density is calculated for each frequency. This is done by flood-filling an \textsc{Aegean}-detected island down to a level four times the local RMS flux density, and integrating.}
\item{For those ($\approx5\sigma$) sources which were not easily fit by a Gaussian, or by a flood-fill, at 120 and 150\,MHz, the position centroid from the source fit to the 180\,MHz data was used to perform a `forced' measurement on that pixel position in the lower-frequency data.}
\end{itemize}
The flux densities at each frequency are then fit in the usual way to produce a single spectral index for the 180-MHz source components.

\begin{figure*}[p]
    \centering
    \begin{subfigure}[b]{0.90\textwidth}
    \centering
                \includegraphics[angle=0,width=0.90\textwidth]{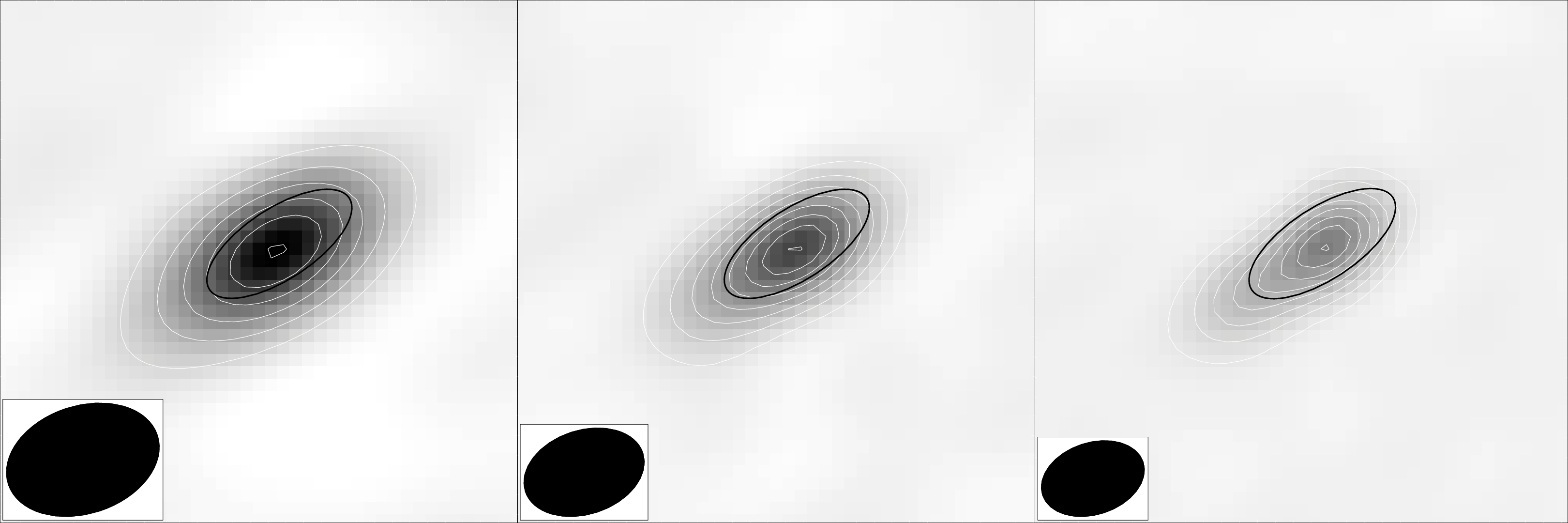}
    \caption{\label{fig:cross-match-extended}}
    \end{subfigure}
    \begin{subfigure}[b]{0.90\textwidth}
    \centering
                \includegraphics[angle=0,width=0.90\textwidth]{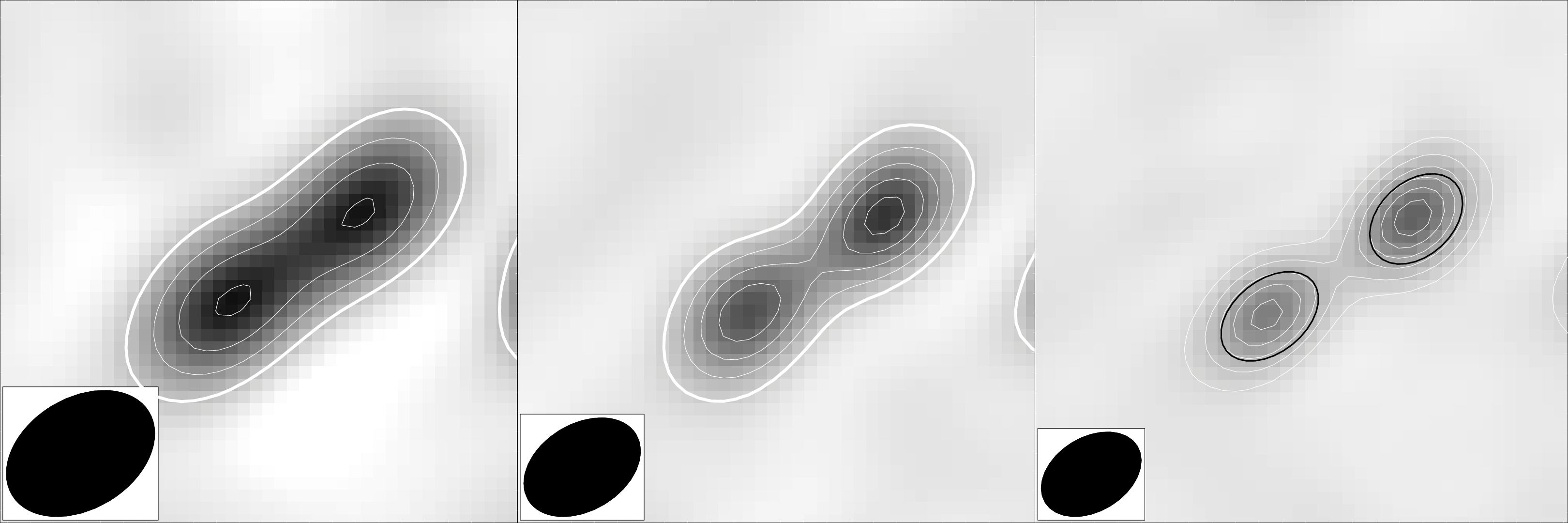}
    \caption{\label{fig:island-flux}}
    \end{subfigure}
    \begin{subfigure}[b]{0.90\textwidth}
    \centering
                \includegraphics[angle=0,width=0.90\textwidth]{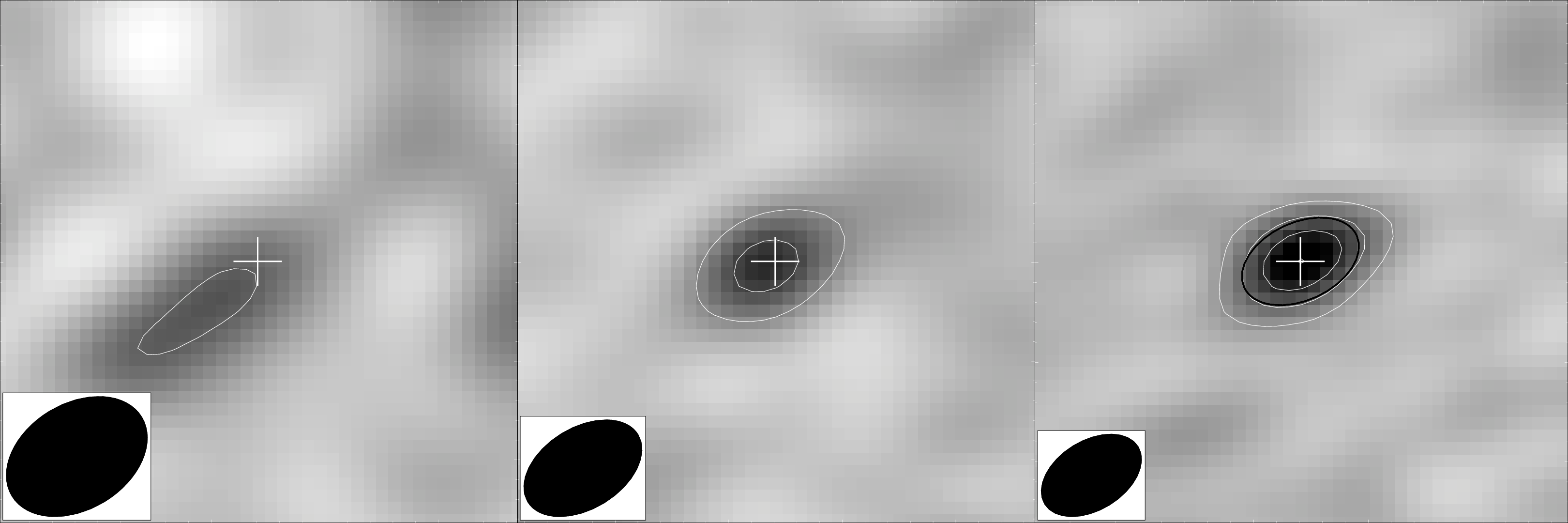}
    \caption{\label{fig:forced-measurement}}
    \end{subfigure}
    \caption{Examples of measuring extended emission; top: cross-matching extended components; middle: integrating a flux density where the source is comprised of multiple components; bottom: forcing measurements on faint sources. In each case, the 180\,MHz image is shown on the left, the 150\,MHz image in the middle, and the 119\,MHz image on the right. All images are square and measure 21\arcmin on each side. The colourscales are linear and the same for each source; the minimum is always $-0.2$Jy\,beam$^{-1}$ and the maxima are 3, 1.5 and 0.4\,Jy\,beam$^{-1}$, from top to bottom. The white contours are linear at $5\sigma$ levels, starting at $4\sigma$; left-to-right, top-to-bottom, the local RMS $\sigma=100$, 60, 42, 71, 47, 30, 54, 30, and 23\,mJy\,beam$^{-1}$. The effective PSF is shown as a filled ellipse in the bottom-left corner. (a) The source is detected as a single extended object at all frequencies, so the \textsc{Aegean} component fits are cross-matched, and the integrated flux densities fitted to produce $S_\mathrm{180\,MHz}$ and $\alpha$ as for unresolved sources. (b) For multiple-component sources where \textsc{Aegean} was unable to detect separate components at the lower frequencies, the components detected at 180\,MHz are used to set up peaks from which the 150\, and 119\,MHz images are flood-filled down to a level of $4\sigma$, shown as a thicker contour. The integrated flux densities of the measurements at all three frequencies are used to fit a spectral index for all of the components resolved and reported at 180\,MHz. (c) For those sources which are not well-fit by a Gaussian, and are very close to $5\sigma$ at the lower frequencies, we force measurements at the position (indicated by a cross) of the 180\,MHz centroid (black ellipse) on the 150 and 120\,MHz mosaics. All three flux density measurements are used to fit a spectral index for the components resolved and reported at 180\,MHz.}
    \label{fig:extended_sources}
\end{figure*}

Examples of each method are shown in \fig~\ref{fig:extended_sources}.
The fitted spectral indices and 180\,MHz flux densities for the components of resolved sources are reported in the catalogue in the same way as for the unresolved sources.
As shown in \tab~\ref{tab:catalogue-extract}, a column in the table indicates whether the spectral index was determined by integrating an extended source, or performing a forced measurement.

Error measurements for the flux densities of these extended sources were identical to those discussed in \sect~\ref{sect:errors}, with the fitting error combined in quadrature with the RMS noise.
It should be noted, however, that the changing resolution of the instrument at different frequencies, combined with the intrinsic complexity of these sources, mean that the automatically derived spectral indices should be treated with caution. 
Postage stamps of the extended sources are available on request
should the reader wish to perform their own measurements.

\section{Source catalogue}
\label{sec:sourcecat}
The MWACS catalogue comprises {\nsrc} sources and is available via Vizier\footnote{http://vizier.u-strasbg.fr/}, and in the electronic edition of this journal. We report the positions of sources, the 180\,MHz integrated flux density and spectral index, the Gaussian fit parameters (major axis, minor axis, and position angle) as well as errors on all of the above parameters. Flags indicate whether a source was confused or extended and thus required special treatment (see \sect~\ref{sect:multi-comps}.) The estimated synthesised beam at that position is also provided. Postage stamps are available on request.
\begin{sidewaystable*}[p]
\footnotesize
\centering
	\caption{Source Catalogue (only the first 15 sources are shown) Table columns are defined as follows: 
(1) IAU name hhmm.m+ddmm.
(2) RA.
(3) Error on RA.
(4) Dec.
(5) Error on Dec.
(6) Integrated Flux density at 180\,MHz.
(7) Error on Integrated flux density.
(8) Spectral index ($\alpha$ where $S_{\nu}\propto\nu^{\alpha}$).
(9) Error on spectral index.
(10) Major axis of source.
(11) Error on major axis.
(12) Minor axis of source. 
(13) Error on minor axis. 
(14) Position angle of source.
(15) Error on position angle of source.
(NB 10, 12 \& 14, are as measured in the 180\,MHz maps and include convolution with the synthesised beam).
(16, 17, 18) Major axis, minor axis and position angle of the PSF at location of source.
(19) Source index within Dec\,$-27$ field.
(20) Source index within Dec\,$-47$ field.
(21) For sources where a multi-component fit was required: the index of the component
(NB sources in the overlap region are identified by having non-null values for both 19\&20. 19, 20 \& 21 can be used to cross-match multiple components. All components will share a value for columns 19 and/or 20 with each component source having a unique value for 21).
(22) Type of spectral fit used to determine source spectral index; 0 indicates source fitting at all three frequencies (\fig~\ref{fig:cross-match-extended}); 1 indicates floodfill (\fig~\ref{fig:island-flux}); 2 indicates a forced measurement (\fig~\ref{fig:forced-measurement}).
}\label{tab:catalogue-extract}
\begin{tabular}{lrrrrrrrrrr}
\hline
  \multicolumn{1}{c}{Name} & 
  \multicolumn{1}{c}{RAJ2000} & 
  \multicolumn{1}{c}{e\_RAJ2000} & 
  \multicolumn{1}{c}{DEJ2000} & 
  \multicolumn{1}{c}{e\_DEJ2000} & 
  \multicolumn{1}{c}{S180} & 
  \multicolumn{1}{c}{e\_S180} & 
  \multicolumn{1}{c}{SpIndex} & 
  \multicolumn{1}{c}{e\_SpIndex} & 
  \multicolumn{1}{c}{MajAxis} & 
  \multicolumn{1}{c}{e\_MajAxis} \\ 
  \multicolumn{1}{c}{} &
  \multicolumn{1}{c}{deg} &
  \multicolumn{1}{c}{arcmin} &
  \multicolumn{1}{c}{deg} &
  \multicolumn{1}{c}{arcmin} &
  \multicolumn{1}{c}{Jy} &
  \multicolumn{1}{c}{Jy} &
  \multicolumn{1}{c}{} &
  \multicolumn{1}{c}{} &
  \multicolumn{1}{c}{arcmin} &
  \multicolumn{1}{c}{arcmin} \\
  \multicolumn{1}{c}{(1)} &
  \multicolumn{1}{c}{(2)} &
  \multicolumn{1}{c}{(3)} &
  \multicolumn{1}{c}{(4)} &
  \multicolumn{1}{c}{(5)} &
  \multicolumn{1}{c}{(6)} &
  \multicolumn{1}{c}{(7)} &
  \multicolumn{1}{c}{(8)} &
  \multicolumn{1}{c}{(9)} &
  \multicolumn{1}{c}{(10)} &
  \multicolumn{1}{c}{(11)} \\
\hline
MWACSJ0000.0-1704	&0.003	&0.5	&-17.070	&0.4	&0.66	&0.09	&-1.0	&0.4	&4.5	&0.4	\\
MWACSJ0000.1-2824	&0.016	&0.5	&-28.406	&0.5	&0.32	&0.04	&-1.1	&0.4	&4.2	&0.4	\\
MWACSJ0000.1-4617	&0.028	&0.6	&-46.297	&0.5	&0.50	&0.07	&-0.8	&0.4	&5.4	&0.5	\\
MWACSJ0000.1-4910	&0.031	&0.4	&-49.182	&0.4	&0.78	&0.09	&-1.5	&0.2	&4.3	&0.2	\\
MWACSJ0000.1-5234	&0.032	&0.6	&-52.577	&0.6	&0.21	&0.04	&-1.7	&0.6	&4.2	&0.6	\\
MWACSJ0000.2-4333	&0.042	&0.5	&-43.553	&0.5	&0.30	&0.04	&-1.4	&0.4	&4.3	&0.5	\\
MWACSJ0000.3-2450	&0.080	&0.7	&-24.844	&0.5	&0.27	&0.04	&-0.4	&0.7	&5.0	&0.7	\\
MWACSJ0000.3-2724	&0.076	&0.6	&-27.414	&0.5	&0.27	&0.04	&-6.1	&0.3	&4.8	&0.6	\\
MWACSJ0000.3-3410	&0.067	&0.5	&-34.174	&0.4	&0.63	&0.08	&-1.1	&0.3	&4.6	&0.4	\\
MWACSJ0000.4-3024	&0.091	&0.7	&-30.409	&0.6	&0.20	&0.04	&-1.2	&0.8	&4.0	&0.6	\\
MWACSJ0000.4-3822	&0.107	&0.4	&-38.383	&0.4	&0.72	&0.09	&-0.7	&0.3	&4.3	&0.3	\\
MWACSJ0000.4-4721	&0.105	&0.4	&-47.363	&0.3	&1.11	&0.12	&-0.7	&0.2	&4.3	&0.2	\\
MWACSJ0000.4-5635	&0.107	&0.5	&-56.589	&0.5	&0.53	&0.07	&-1.4	&0.4	&4.7	&0.5	\\
MWACSJ0000.5-3320	&0.119	&0.5	&-33.337	&0.4	&0.68	&0.09	&-0.6	&0.3	&4.7	&0.4	\\
MWACSJ0000.5-3452	&0.125	&0.5	&-34.875	&0.4	&0.90	&0.11	&-1.4	&0.3	&4.5	&0.4	\\
\hline\end{tabular}
%
%
\begin{tabular}{rrrrrrrrrcr}
\hline
  \multicolumn{1}{c}{MinAxis} & 
  \multicolumn{1}{c}{e\_MinAxis} & 
  \multicolumn{1}{c}{PA} & 
  \multicolumn{1}{c}{e\_PA} & 
  \multicolumn{1}{c}{MajAxisBeam} & 
  \multicolumn{1}{c}{MinAxisBeam} & 
  \multicolumn{1}{c}{PABeam} & 
  \multicolumn{1}{c}{ID\_C102} & 
  \multicolumn{1}{c}{ID\_C103} & 
  \multicolumn{1}{c}{Component\_ID} & 
  \multicolumn{1}{c}{Fit} \\ 
  \multicolumn{1}{c}{arcmin} &
  \multicolumn{1}{c}{arcmin} &
  \multicolumn{1}{c}{deg} &
  \multicolumn{1}{c}{deg} &
  \multicolumn{1}{c}{arcmin} &
  \multicolumn{1}{c}{arcmin} &
  \multicolumn{1}{c}{deg} &
  \multicolumn{1}{c}{} &
  \multicolumn{1}{c}{} &
  \multicolumn{1}{c}{} &
  \multicolumn{1}{c}{} \\
  \multicolumn{1}{c}{(12)} &
  \multicolumn{1}{c}{(13)} &
  \multicolumn{1}{c}{(14)} &
  \multicolumn{1}{c}{(15)} &
  \multicolumn{1}{c}{(16)} &
  \multicolumn{1}{c}{(17)} &
  \multicolumn{1}{c}{(18)} &
  \multicolumn{1}{c}{(19)} &
  \multicolumn{1}{c}{(20)} &
  \multicolumn{1}{c}{(21)} &
  \multicolumn{1}{c}{(22)} \\
\hline
2.9	&0.3	&-67	&27	&4.4	&2.8	&-63	&4760	&\dots	&	&0	\\
3.0	&0.3	&-56	&32	&4.4	&2.8	&-64	&4587	&\dots	&	&0	\\
3.3	&0.3	&-60	&25	&4.3	&2.9	&-51	&\dots	&4229	&\dots	&0	\\
3.1	&0.2	&-45	&24	&4.3	&3.0	&-50	&\dots	&1620	&\dots	&0	\\
3.0	&0.4	&-53	&36	&4.4	&3.0	&-48	&\dots	&6683	&\dots	&0	\\
3.1	&0.3	&-59	&32	&4.3	&2.9	&-52	&\dots	&4870	&\dots	&0	\\
2.9	&0.4	&-74	&29	&4.4	&2.8	&-64	&7128	&\dots	&	&0	\\
2.7	&0.3	&-56	&26	&4.4	&2.8	&-64	&5995	&\dots	&1	&1	\\
3.0	&0.3	&-66	&26	&4.4	&2.8	&-63	&4910	&\dots	&	&0	\\
2.8	&0.4	&-67	&38	&4.4	&2.8	&-63	&6951	&\dots	&	&0	\\
3.0	&0.2	&-57	&26	&4.2	&2.9	&-54	&\dots	&3425	&\dots	&0	\\
3.0	&0.2	&-53	&21	&4.3	&2.9	&-50	&\dots	&1358	&\dots	&0	\\
3.1	&0.3	&-41	&28	&4.5	&3.1	&-45	&\dots	&4385	&\dots	&0	\\
3.0	&0.3	&-64	&24	&4.4	&2.8	&-63	&3834	&\dots	&	&0	\\
2.9	&0.2	&-68	&24	&4.4	&2.8	&-63	&3443	&\dots	&	&0	\\
\hline\end{tabular}
\end{sidewaystable*}

\subsection{Catalogue reliability and completeness}
Recall that all of the source-finding for this survey was performed on the highest-frequency mosaic.
For point sources, after source identification, the source was fitted at all other frequencies.
If any of these fits failed, the source would be treated as resolved; see \sect~\ref{sect:multi-comps} for details.
We searched for and identified ten spurious sources in the sidelobes of bright sources by searching the areas around ($\approx100$) sources of $S>10$\,Jy, comparing images at all frequencies, and removed these from the catalogue; the faintest source which produced a spurious sidelobe source was 14\,Jy. In typical circumstances away from bright sources, we require that all sources are detectable at $5\sigma$ at 180\,MHz and well-fit by a Gaussian in the other bands, so with a Gaussian random background, the chance of any source being spurious is vanishingly small. Our noise is somewhat non-Gaussian due to the presence of undeconvolved source sidelobes, but in the general field, these sidelobes are faint due to our good $u,v$-coverage, which suppresses sidelobes down to a $10\%$ level even before deconvolution; and the positions of change sidelobe alignments would vary with frequency and thus be unlikely to be detected in every channel.
Therefore we can have an extremely high confidence that all of the sources in the catalogue are real.

The varying detection threshold affects the catalogue completeness for two reasons.
More obviously, the variations in the local RMS cause a varying absolute detection limit across the field.
In addition, the noise in the map introduces a non-zero probability that a source whose intrinsic flux is $>5\sigma$, has a measured flux $<5\sigma$ due to instrumental effects (and vice versa).
We therefore present the noise properties of our maps in some detail.

\subsubsection{Variations in RMS sensitivity}\label{sect:noise}

\begin{figure}
\centering
		\includegraphics[width=0.5\textwidth]{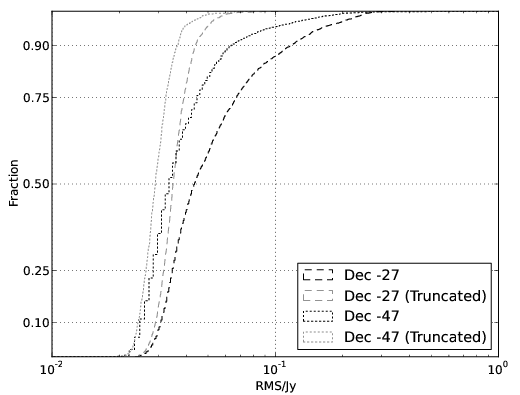}
		\caption{Cumulative distribution functions of the RMS at 180\,MHz for those areas of the map outside of the overlap region for the Dec\,$-27$ scan (black dashed line) and Dec\,$-47$ scan (black dotted line). A subset of data truncated in RA and Dec range is also shown for both scans: Dec\,$-27$ (grey dashed line: $22^{\mathrm{h}}00^{\mathrm{m}}<\alpha<07^{\mathrm{h}}30^{\mathrm{m}}$, $-30\arcdeg<\delta<-20\arcdeg$); Dec\,$-47$ (grey dotted line:  $21^{\mathrm{h}}15^{\mathrm{m}}<\alpha<06^{\mathrm{h}}40^{\mathrm{m}}$, $-42\arcdeg<\delta<-52\arcdeg$)}
		\label{fig:rms_cdf}
%
\end{figure}

\Fig~\ref{fig:rmsmap} shows maps of the local RMS of the mosaics rescaled to take into account absolute calibration (with XX and YY combined together for the Dec\,$-47$ scan).
This information is summarised in \fig~\ref{fig:rms_cdf}.
\begin{figure}
	\centering
	\begin{subfigure}{0.5\textwidth}
                \centering
                \includegraphics[width=\textwidth]{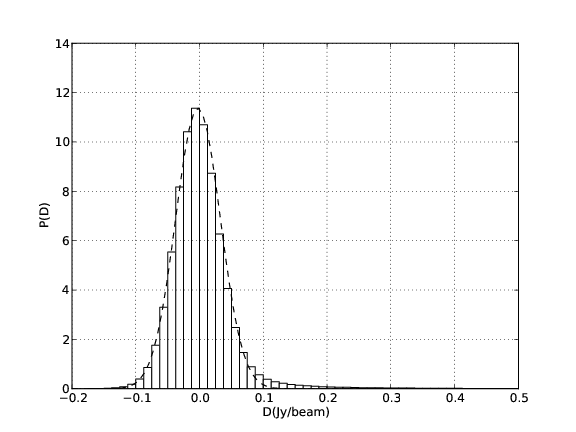}
                \label{fig:c102_hist}
	\end{subfigure}
	\begin{subfigure}{0.5\textwidth}
                \centering
                \includegraphics[width=\textwidth]{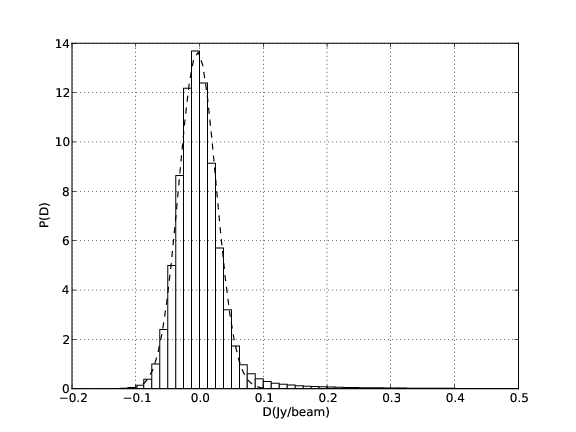}
                \label{fig:c103_hist}
	\end{subfigure}
	\caption{Histogram of the pixel brightnesses of the 180\,MHz maps (P(D)). RA and Dec ranges are $22^{\mathrm{h}}00^{\mathrm{m}}<\alpha<07^{\mathrm{h}}30^{\mathrm{m}}$, $-30\arcdeg<\delta<-20\arcdeg$ (approx 1300 square degrees) for the Dec\,$-27$ mosaic (top) and $21^{\mathrm{h}}15^{\mathrm{m}}<\alpha<06^{\mathrm{h}}40^{\mathrm{m}}$, $-42\arcdeg<\delta<-52\arcdeg$ (approx 1000 square degrees) for the Dec\,$-47$ mosaic (bottom). The RMS for each pixel distribution (determined via the semi-interhexile range) is 0.035\,Jy\,beam$^{-1}$ and 0.029\,Jy\,beam$^{-1}$ respectively. The dashed line is a gaussian curve with this RMS, centred on the median flux density ($-2.4$\,mJy\,beam$^{-1}$ and $-3.2$\,mJy\,beam$^{-1}$ respectively).}
    \label{fig:hist}
\end{figure}
This figure shows that the extremely high RMS regions associated with bright extended sources, particularly near the Galactic plane, account for a small proportion of survey area.
Overall, the sensitivity of the survey varies about the median of $\sim$40\,mJy by less than a factor of two over 80\% of the survey area, a variation in sensitivity comparable with the VLSS \citep[c.f.][\fig~7]{2007AJ....134.1245C}.

The RMS in the Dec\,$-27$ and Dec\,$-47$ mosaics are reasonably comparable, though the Dec\,$-47$ mosaic has a somewhat lower RMS, particularly for regions below the median.
We ascribe this to a better phase calibration for the Dec\,$-47$ data, as a brighter phase calibrator was used.
The RMS becomes only marginally worse towards the edge of the primary beam in the Dec\,$-47$ mosaic, however it becomes significantly worse towards the edge of the primary beam in the Dec\,$-27$ mosaic.
This is because the sky rotates more quickly through the primary beam closer to the celestial equator.

\Fig~\ref{fig:hist} shows the histogram of flux-density values across a significant proportion of each mosaic.
When calculated over a wide area, the RMS for the two maps agrees to $\sim$10\%; 90\% of the surveyed area has an RMS lower than 50\,mJy, except for the very Northern edge.

\subsection{Fornax~A}

Fornax~A is a nearby ($z\approx0.006$) radio galaxy, situated in the Fornax cluster of galaxies. It comprises two large lobes extending NNW and SSE 20\arcmin ($\approx200$\,kpc) from the lenticular galaxy NGC~1316, at RA$=03^{\mathrm{h}}22^{\mathrm{m}}41.789^{\mathrm{s}}$ Dec$=-37^{\circ}12\arcmin29\farcs52$. On scales $\gtrsim20$\arcmin, structure is not sampled by MWACS, so we cannot measure the integrated flux density for this object, only set lower limits. This under-sampling also causes deconvolution errors across our mosaics, resulting in increased noise levels around the Fornax~A region.
\begin{figure}
\centering
\includegraphics[scale=1, angle=-90, width=0.5\textwidth]{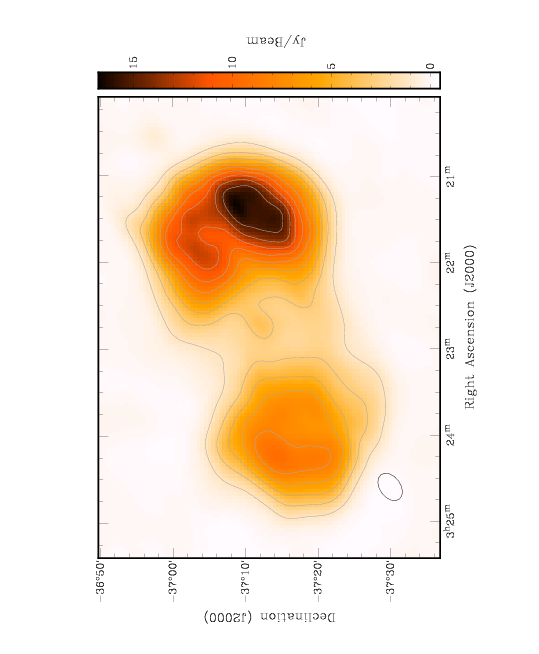}
\caption{Fornax~A at 180\,MHz; contours are linear at 10\% levels from the peak brightness of 16.75\,Jy\,beam$^{-1}$. The synthesised beam is shown as an open ellipse at the bottom-left of the image, with dimensions $4\farcm22\times2\farcm85$, position angle $-53\fdg8$. The RMS of this image is 92\,mJy.}\label{fig:fornax}
\end{figure}

Using \textsc{BlobCat} to flood-fill the entire region down to a $4\sigma$ level and measure the integrated flux density, we set lower limits across the entirety of Fornax~A at $786\pm  39$, $668\pm  33$, and $514\pm  26$\,Jy at 120, 150 and 180\,MHz, respectively; these are consistent with the values measured by \citet{2013ApJ...771..105B} and \citet{2014McKinley}. An image of the radio galaxy at the highest MWACS resolution is shown in \fig~\ref{fig:fornax}.
\section{Discussion and conclusions}
\label{sec:conclusions}
We have performed a survey of approximately {\survarea} square degrees of the southern sky during the MWA's commissioning period in 2012~October: the MWA Commissioning Survey (MWACS). The observations were made in meridian drift scan mode at two Dec settings ($-26\fdg7$ and $-47\fdg5$) centred on three frequencies: 119, 150 and 180\,MHz, resulting in an effective frequency coverage of 104 to 196 MHz.
Because the data were taken as drift scans, and because the MWA's primary beam X and Y polarisation responses are different when off-zenith, some unconventional data processing was required.
Six separate mosaic images were formed, one for each frequency and Dec covering the entire RA range of the drift scan.
By taking advantage of the instantaneous (near) co-planarity of the array and good instantaneous $u,v$-coverage (including large fractional bandwidth), the data were processed as continuum multi-frequency synthesis images (at the three central frequencies) by forming a mosaic of weighted regridded snapshot images.

The source 3C32 was used to set the absolute flux density scale. Stokes~I images at 119, 150 and 180\,MHz were formed by adding the primary-beam weighted XX and YY mosaics. The absolute flux density was bootstrapped from the Dec\,$-27$ mosaic to the Dec\,$-47$ mosaic using sources in the overlapping five degrees of Declination.
We found the primary beam response at Dec\,$-47$ to be sufficiently different between the XX and YY polarisations that an independent scaling factor was used for each before being combined to Stokes~I.

The data do not have sufficient short baselines to recover large-scale structure, so only compact sources were extracted from the mosaics. We first identified sources as islands of $>5\sigma$ flux density, then separated sources into single-component (unresolved) and multi-component (resolved, and/or multiple nearby components) categories, and measured their flux densities at each frequency, appropriately.
We present the result of the MWACS a single catalogue comprised of 180-MHz flux densities and spectral indices calculated from the three individual measured frequencies, for {\nsrc} sources.

The sensitivity of MWACS, as measured by local image RMS, is mostly consistent around 40\,mJy\,beam$^{-1}$ ($1\sigma$) across the mosaics but degrades near bright sources and near the edges of the fields.
All sources found in MWACS at 180\,MHz were also found in the two lower frequency mosaics and had reasonable spectra given their S/N.
We therefore consider detections to be reliable subject to the usual considerations on confusion and resolution.
Users are advised, however, that non-detections can only be used to set an upper limit to flux density in the context of the local RMS.
Regions within a few degrees of the edge of the image or bright sources will have locally higher RMS.

This commissioning survey demonstrates the impressive survey capabilities of the MWA and the feasibility of processing MWA drift scan datasets using fairly conventional radio astronomy tools with simple mosaicking techniques.
The full MWA has near-complete snapshot $u,v$-coverage between approximately 5~and~500 wavelengths (as multi-frequency synthesis images) and with 128 fully cross-correlated tiles, its quality of calibration is higher than our combined commissioning sub-arrays.
We therefore expect the GLEAM survey to be confusion limited in Stokes~I, to be able to capture large-scale structures (up to $\sim$10\arcdeg) and to have much reduced imaging artifacts from better calibration and deconvolution.

The progression of the MWA from 32T prototype (circa 2009--2011) through the commissioning array (late 2012 through mid-2013) and into operations (mid-2013 on) has shown a steady improvement in area and depth of sky coverage and of our understanding of the instrument and data.
The GLEAM survey, currently underway, has derived great benefit from the accumulated expertise of the 32T and commissioning phases and will likewise provide increased understanding of the instrument, especially the primary beam and resulting polarisation performance of the MWA, and data processing methods.
Looking forward to SKA-low, we expect the expertise gained from processing the large GLEAM dataset on world-class supercomputers to be provide equally valuable lessons when SKA-low Phase\,1 is complete.
\section*{Acknowledgements}
This scientific work makes use of the Murchison Radio-astronomy Observatory, operated by CSIRO. We acknowledge the Wajarri Yamatji people as the traditional owners of the Observatory site. Support for the MWA comes from the U.S. National Science Foundation (grants AST-0457585, PHY-0835713, CAREER-0847753, and AST-0908884), the Australian Research Council (LIEF grants LE0775621 and LE0882938), the U.S. Air Force Office of Scientific Research (grant FA9550-0510247), and the Centre for All-sky Astrophysics (an Australian Research Council Centre of Excellence funded by grant CE110001020). Support is also provided by the Smithsonian Astrophysical Observatory, the MIT School of Science, the Raman Research Institute, the Australian National University, and the Victoria University of Wellington (via grant MED-E1799 from the New Zealand Ministry of Economic Development and an IBM Shared University Research Grant). The Australian Federal government provides additional support via the Commonwealth Scientific and Industrial Research Organisation (CSIRO), National Collaborative Research Infrastructure Strategy, Education Investment Fund, and the Australia India Strategic Research Fund, and Astronomy Australia Limited, under contract to Curtin University. We acknowledge the iVEC Petabyte Data Store, the Initiative in Innovative Computing and the CUDA Center for Excellence sponsored by NVIDIA at Harvard University, and the International Centre for Radio Astronomy Research (ICRAR), a Joint Venture of Curtin University and The University of Western Australia, funded by the Western Australian State government. This research has made use of the NASA/IPAC Extragalactic Database (NED) which is operated by the Jet Propulsion Laboratory, California Institute of Technology, under contract with the National Aeronautics and Space Administration.

\newcommand{\pasa}{PASA}
\newcommand{\aj}{AJ}
\newcommand{\apj}{ApJ}
\newcommand{\apjs}{ApJS}
\newcommand{\apjl}{ApJL}
\newcommand{\aap}{A{\&}A}
\newcommand{\aaps}{A{\&}AS}
\newcommand{\mnras}{MNRAS}
\newcommand{\araa}{ARAA}
\newcommand{\pasp}{PASP}
\bibliographystyle{apj}
\bibliography{refs}

\end{document}